\documentclass[12pt]{article}

\usepackage{amsmath}
\usepackage{amssymb}
\usepackage{slashed}
\usepackage[numbers,sort&compress]{natbib}
\usepackage{hyperref}
\usepackage{cleveref}
\crefname{equation}{Eq.}{Eqs.}
\crefname{figure}{Fig.}{Figs.}
\crefname{table}{Table}{Tables}
\crefname{section}{Section}{Sections}

\usepackage[letterpaper,margin=1in,bottom=1in]{geometry}
\usepackage{float} 
\usepackage{parskip} 
\usepackage{tabulary} 
\usepackage{color} 
\usepackage{soul} 
\usepackage{subfigure}
\usepackage{graphicx}
\usepackage[section]{placeins} 

\def\lhc2{LHC~Run~II}
\def\nsu{nonuniversal~supergravity~models~}

\usepackage{cleveref}

\newcommand{\code}[1]{\texttt{#1}}

\def\.4{\vspace{-.5cm}}
\newcommand{\ifb}{~\textrm{fb}^{-1}}
\newcommand{\iab}{~\textrm{ab}^{-1}}

\def\beq{\begin{equation}}
\def\be{\begin{equation}}
\def\beqn{\begin{eqnarray}}
\def\ee{\end{equation}}
\def\eeq{\end{equation}}
\def\eeqn{\end{eqnarray}}

\author{
Amin Aboubrahim\footnote{Email: a.abouibrahim@northeastern.edu}~\ and 
Pran Nath\footnote{Email: p.nath@northeastern.edu}\\~\\
Department of Physics, Northeastern University,
Boston, MA 02115-5000, USA
}

\title{Supersymmetry at  a 28 TeV  hadron collider: HE-LHC}

\begin{document}
\maketitle
\date

\textbf{Abstract: } 
The discovery of the Higgs boson at $\sim 125$ GeV indicates that the scale of weak scale supersymmetry is higher than what was perceived in the pre-Higgs boson 
discovery era and lies in the several TeV region. This makes the discovery of supersymmetry more challenging and argues for hadron colliders beyond LHC at
$\sqrt s=14$ TeV.  The Future Circular Collider (FCC) study at CERN is considering a 100 TeV collider to be installed in a 100 km tunnel in the Lake Geneva basin.
Another 100 km collider being considered  in China is the Super proton-proton Collider (SppC).  A third possibility recently 
proposed is the High-Energy LHC (HE-LHC) which would use the existing CERN tunnel but achieve a center-of-mass energy of 28 TeV by using 
FCC magnet technology at significantly higher luminosity than at the High Luminosity LHC (HL-LHC).
 In this work we investigate the potential of HE-LHC for the discovery of supersymmetry.
We study a class of supergravity unified models under the Higgs boson  mass and  the dark matter relic density constraints and compare the analysis
with the potential reach of the HL-LHC. A set of benchmarks are presented which are beyond the discovery potential of HL-LHC but are discoverable 
at HE-LHC. For comparison, we study model points at HE-LHC  which are also discoverable at HL-LHC. For these model points, it is found
that their discovery would require a HL-LHC run between 5-8 years while the same  parameter points can be discovered in a period of few weeks to
$\sim 1.5$ yr at HE-LHC running at its optimal luminosity of $2.5\times 10^{35}$ cm$^{-2}$ s$^{-1}$.
The analysis indicates that the HE-LHC possibility should be seriously pursued as it would significantly increase the
discovery reach for supersymmetry  beyond that of HL-LHC and decrease the run period for discovery. 

\newpage

\section{Introduction}\label{sec:intro}
The discovery of the Higgs boson~\cite{Englert:1964et, Higgs:1964pj, Guralnik:1964eu}  mass at $\sim 125$ GeV ~\cite{Chatrchyan:2012ufa, Aad:2012tfa}
has put stringent constraints on the scale of weak scale supersymmetry.
Thus within supersymmetry and supergravity unified theories a Higgs boson mass of $\sim 125$ GeV requires a very significant loop correction
which points to the scale of weak scale supersymmetry lying in the several TeV region~\cite{Akula:2011aa,Arbey:2012dq,susy-higgs,Baer:2015fsa}.
As a result, the observation of the Higgs boson mass at 
$\sim 125$ GeV makes the discovery of supersymmetry more difficult. This difficulty arises on two fronts.  First, 
the large scale of weak scale supersymmetry implies that the average mass of the sparticles, specifically of sfermions,  is significantly higher than what was 
thought in the pre-Higgs boson discovery era. This leads to a suppression in the production of sparticles at colliders. Second, in high scale unified models
such as supergravity grand unified models (SUGRA)~\cite{msugra} (for a review see~\cite{Nath:2016qzm})
with R-parity conservation, the  satisfaction of the relic density consistent with WMAP~\cite{Larson:2010gs} and the PLANCK~\cite{Ade:2015xua} experimental 
data
 requires coannihilation~\cite{Griest:1990kh} which means that the next to lightest supersymmetric 
particle (NLSP) lies close to the lightest supersymmetric particle (LSP).  The close proximity of the NLSP to the LSP means that the decay
of the NLSP to the LSP will result in light detectable final states, i.e., leptons and jets, making the detection of supersymmetry more difficult.  

In view of the above, the non-observation of supersymmetry thus far is not surprising. In fact, as argued recently, the case for supersymmetry
 is stronger after the Higgs boson discovery~\cite{Nath:2017lma} and we discuss briefly the underlying 
  reasons for the pursuit of supersymmetry.  Thus one of the attractive features of
   supersymmetry is the resolution of
  the large hierarchy problem related to the quadratic divergence of the loop correction to the Higgs boson mass by quark loops and its
  cancellation by squark loops. This situation is very much reminiscent of the cancellation of the up quark contribution by the charm quark
  contribution in resolving the flavor changing neutral current problem which lead to the discovery of the charm quark.  In that case 
  a natural   cancellation occurred up to one part in $10^9$ while for the Higgs boson case a natural cancellation occurs up to one
  part in $10^{28}$  in order to  cancel the quadratic divergence in the  Higgs boson mass square.
    Thus the cancellation for the Higgs boson case is even more compelling than for the flavor changing neutral current case. 
    Aside from that, a heavy weak scale of supersymmetry resolves some problems specific to supersymmetry. Thus supersymmetry brings with it 
 new CP violating phases which can generate very large EDMs for the quarks and the leptons which are in violation of experiment 
 if the squark and slepton masses are in the sub TeV region. A solution to this problem requires fine tuning, or a cancellation mechanism~\cite{Ibrahim:1997gj,Ibrahim:2007fb}.
 However, if the squark and slepton masses are large, one has a more natural suppression of the EDM consistent with experiment~\cite{Nath:1991dn,Kizukuri:1992nj}. 
 
Another potential problem for a low scale of weak scale supersymmetry concerns proton decay from baryon and lepton number 
violating dimension five operators. For a low scale of weak scale supersymmetry, a suppression of this again requires a
fine tuning but for scale of weak scale supersymmetry lying in the several TeV region this suppression is more easily 
accomplished~\cite{Liu:2013ula,Nath:2006ut}. 
We note in passing that the unification of gauge coupling constants is satisfied to a good degree of accuracy in models with 
scalar masses  lying in the tens of TeV as for the case when the weak scale of supersymmetry lies in the sub TeV region~\cite{Aboubrahim:2017wjl}.

The LHC has four phases which we may label as LHC1-LHC4.  The LHC1 phase at $\sqrt s=7-8$ TeV lead to the discovery of the Higgs boson.  
We are currently in the LHC2 phase where $\sqrt s=13$ TeV  and it will continue till the  end of 2018 and by that time the CMS and  the ATLAS detectors are each expected to 
collect 150 fb$^{-1}$ of integrated luminosity. LHC will then shut down for two years in the period 2019-2020 for 
upgrade to LHC3 which will operate at 14 TeV in the period 2021-2023. In this period each of the detectors will 
collect additional  300 fb$^{-1}$ of data. LHC will then shut down for a major upgrade to high luminosity LHC (HL-LHC or LHC4) for a two
and a half years in the period 2023-2026 and will resume operations in late 2026  and run for an expected 10 year period
till 2036. At the end of this period it is expected that each detector will collect additional data culminating in 3000 fb$^{-1}$.
Beyond LHC4, higher energy $pp$ colliders have been discussed.  These include 
 a 100 TeV hadron collider  at CERN 
 and a  100 TeV  proton-proton collider in China each of which requires a circular ring of about 100 km~\cite{Arkani-Hamed:2015vfh,Mangano:2017tke}.

Recently, a 28 TeV $pp$ collider at CERN has been discussed~\cite{Benedikt:2018ofy,Zimmermann:2018koi,HE-LHC-1,HE-LHC-2,cern-report} as a third 
possibility for a hadron collider beyond the LHC
which has the virtue that it could be 
 built using the existing ring at CERN
by installing 16 T 
superconducting magnets  using FCC technology
capable of enhancing the center-of-mass energy of the collider to 28 TeV. Further, HE-LHC will operate at a luminosity of $2.5\times 10^{35}$ cm$^{-2}$s$^{-1}$ 
and collect 10-12 ab$^{-1}$ of data. 
This set up necessarily means  that a larger part of the parameter space of supersymmetric models beyond the reach of the 14 TeV collider will be probed.
Also, supersymmetric particles that could be discovered at the HL-LHC may be discoverable at 28 TeV at a much lower integrated luminosity. 
In this work we investigate supersymmetry signatures at LHC-28 (or HE-LHC) and compare the integrated luminosity necessary for a 5$\sigma$ discovery of a set of supergravity benchmark points with what one would obtain at LHC-14. The analysis is done under the constraints of the Higgs boson mass at 125 $\pm$ 2 GeV and the relic density 
constraint on neutralino dark matter of
 $\Omega_{\tilde \chi_1^0} h^{2}<0.128$. For a bino-like LSP, satisfaction of  the relic density constraint  requires coannihilation. Specifically for the set of benchmarks considered, the chargino is the NLSP and one has chargino coannihilation in cases where the LSP is bino-like. Here we use \nsu with nonuniversalities in the gaugino (and Higgs) sector to investigate a range of neutralino, chargino and gluino masses that are discoverable at the HE-LHC.\\
The outline of the rest of the paper is as follows: In section \ref{sec2} we discuss the  SUGRA models and the benchmarks investigated in this work. 
These benchmarks are listed in Table~\ref{tab2} and they satisfy all the desired constraints.  In section \ref{sec3} we discuss the prominent discovery
channels used to investigate the discovery of the benchmarks at $\sqrt s=14$ and $\sqrt s=28$ TeV. Here we first discuss the various codes used
in the analysis. It is found that the most prominent channels for discovery include single lepton, two lepton, and three lepton channels along with jets.
Details of these analyses are given in sections \ref{sec3.1}, \ref{sec3.2}, and \ref{sec3.3}.
Thus in section \ref{sec3.1} an analysis of the benchmarks using a single lepton and jets in the final state is investigated; in section \ref{sec3.2} an analysis using two leptons and
jets  in the final states is discussed and in section \ref{sec3.3}, an analysis is given using 3 leptons and jets in the final state.  An estimate of uncertainties is given in 
section \ref{sec-uncern}.
A discussion of dark matter direct detection for the benchmarks of Table~\ref{tab2} is given in section \ref{sec4} while conclusions are given in section \ref{sec5}. The analysis of this work is illustrated by 
several tables and figures which are called at appropriate points in the various sections. 
 
\section{ SUGRA model benchmarks \label{sec2}}
In Table~\ref{tab2} we give a set of benchmark SUGRA models. These models are consistent with the constraints of 
radiative breaking of the electroweak symmetry (for review see~\cite{Ibanez:2007pf}),
 the Higgs boson mass constraint and  the relic density constraint.  In high scale models
the neutralino often turns out to be mostly a bino and thus its annihilation requires the presence of another sparticle in close proximity, i.e., 
coannihilation (for early work see~\cite{Griest:1990kh}).
Coannihilation arises in supergravity models in a variety of ways with universal  as well as with non-universal boundary conditions at 
the grand unification scale, $M_G$, taken to be $2\times 10^{16}$ GeV. The non-universalities include those 
 in the gaugino sector~\cite{Ellis:1985jn,nonuni2,Belyaev:2018vkl},  in the matter sector  and in the Higgs sector~\cite{NU}. 
These boundary conditions lead to a vast 
 landscape of sparticle mass hierarchies~\cite{Feldman:2007zn}.  
 Coannihilation necessarily leads to a partially compressed sparticle spectrum. [Compressed spectra have been investigated in a number of 
 recent works see, e.g., \cite{Kaufman:2015nda, Nath:2016kfp, Aboubrahim:2017aen,Dutta:2015exw,Berggren:2015qua,Berggren:2016qjh,LeCompte:2011fh}. For experimental searches for supersymmetry with  compressed spectra see~\cite{Khachatryan:2016mbu,Khachatryan:2016pxa,MORVAJ:2014opa})].
For the models of Table~\ref{tab2} the NLSP is the light chargino whereby for points (g)-(j) the LSP is bino-like and satisfaction of  the relic density constraint is realized by chargino coannihilation. To achieve chargino coannihilation we need to have non-universal supergravity models with non-universalities in the $SU(2)$ and  the $SU(3)$ sectors. Thus the parameter space of the models is given by 
\begin{align}
 m_0 \, , A_0 \, , m_1 \, , m_2 \, , m_3 \, , \tan\beta \, , \text{sgn}(\mu)\, ,
 \label{msugra}
 \end{align}
where $m_0$ is the universal scalar mass, $A_0$ is the universal trilinear scalar coupling at the grand unification scale, $\tan\beta=\langle H_2\rangle/\langle H_1\rangle$, where $H_2$ gives mass to the up-type quarks and $H_1$ gives mass to the down-type quarks and the leptons, and sgn$(\mu)$ is the sign of the Higgs mixing parameter $\mu$ 
 which enters in the superpotential in the form $\mu H_1 H_2$. In the analysis we consider values of the universal scalar mass which, although high, arise
 quite naturally  on the hyperbolic branch of radiative breaking of the electroweak symmetry. All of the parameter points given in  Table~\ref{tab2} are not
 currently probed at the LHC and are thus not ruled out by experiment~\cite{Aad:2014vma,Aad:2015iea,Aad:2015jqa,Aaboud:2017leg,Aaboud:2017vwy,Aaboud:2017bac,Aaboud:2018jiw}. Table~\ref{tab3} shows qualitatively two types of parameter points. 
 Thus model points (a)-(f) have the neutralino and the chargino masses close to 1 TeV while the gluino is also relatively light. For model points 
 (g)-(j) of Table~\ref{tab2}, 
  we have neutralino and the chargino masses lying below 200 GeV while the gluino is much heavier. In each case  the Higgs boson mass and the relic density constraints are satisfied and a compressed spectrum is obtained. For model points (a)-(f) of Table~\ref{tab2} 
  the compressed spectrum involves 
 $\tilde\chi_1^0, \tilde\chi_2^0, \tilde\chi_1^{\pm}, \tilde g$ 
 while for models points (g)-(j) of Table~\ref{tab2}, 
 it involves $\tilde\chi_1^0, \tilde\chi_2^0, \tilde\chi_1^{\pm}$ (see Fig.~\ref{spectrum}).
 In these models the sfermions are all heavy with the lightest sfermion mass  lying in the several TeV range.  
However, as  argued previously, the  TeV size scalars can be quite natural in SUGRA models since
  the weak scale could be
large and natural on the hyperbolic branch  
of radiative breaking of the electroweak symmetry in SUGRA models~\cite{Chan:1997bi,Feng:1999mn,Chattopadhyay:2003xi,Baer:2003wx,Feldman:2011ud,Akula:2011jx,Ross:2017kjc}. 
We note here that in the analysis of \cite{Chan:1997bi},  the ratio $f=\mu^2/M_Z^2$ was suggested as a criteria of fine tuning. For the model points of Table 3, we see that $\sqrt f$ lies in the range $\sim 13$ (for i) to $\sim  347$ (for point j). As noted 
above the large Higgs mass correction requires that  the weak SUSY scale lie in the several TeV region and the analysis of Table 3 reflects that reality. 
\\

The analysis is performed at the current LHC energy of 14 TeV and at the proposed center-of-mass energy of 28 TeV. For each of the model points defined by Eq.~(\ref{msugra}) at the GUT scale, the RGE's are run down to the electroweak scale to obtain the entire SUSY sparticle spectrum. This is performed using \code{SoftSUSY 4.1.0}~\cite{Allanach:2001kg, Allanach:2016rxd} which determines the Higgs boson mass at the two-loop level. The input parameters of the SUGRA model are given in Table~\ref{tab2} and the results obtained from \code{SoftSUSY} are shown in Table~\ref{tab3} consistent with the Higgs boson mass and the relic density constraints, with the latter calculated using \code{micrOMEGAs 4.3.2}~\cite{Belanger:2014vza}. SUSY Les Houches Accord formatted data files are processed using \code{PySLHA}~\cite{Buckley:2013jua}. It is clear from Table~\ref{tab3} that the mass difference between the chargino and the LSP is small, i.e., $$(m_{\tilde{\chi}^{\pm}_1}-m_{\tilde{\chi}^0_1})\ll m_{\tilde{\chi}^{0}_1},$$ and is essential to drive the relic density to within the observed limits in a parameter space where the LSP is bino-like. This is observed for points (g)-(j), while points (a)-(f) have an LSP which is wino-like and this explains the lower values of the relic density. There is a certain range of LSP-chargino mass gap which can still keep the relic density in check. As mentioned before, the LSP, chargino, stop and gluino masses presented in this analysis are still not excluded by the current LHC analyses. \\
We analyze final states coming from the production and subsequent decay of a second neutralino in association with a chargino on one hand and a gluino pair on the other hand. The leading order (LO) production cross-sections of $\tilde{\chi}_2^{0}\tilde{\chi}_1^{\pm}$ and $\tilde{g}\tilde{g}$ are presented in Table~\ref{tab4} for two LHC center-of-mass energies: 14 and 28 TeV. A plot of the production cross-sections of the ten benchmark points is presented in Fig.~\ref{fig1} where the left panel exhibits the gluino pair production cross-sections and the right panel is for the second neutralino-chargino production, both plotted against different center-of-mass energies. Note that the next-to-leading order and next-to-leading logarithm (NLO+NLL) cross-sections for the gluon pair production reported by the LHC SUSY cross section working group are given at fixed order in QCD. In our calculation, the $\tilde{g}\tilde{g}$ production cross-section is evaluated at LO with hard jets at generator level and then matched with parton shower which is why the cross-sections appear to be way smaller. At fixed order, the LO cross-sections at $\sqrt{s}=14$ TeV for points (a) to (f) would become 0.053, 0.041, 0.022, 0.006, 0.066 and 0.012 pb, respectively. Comparing those values to the NLO+NLL ones, the difference is very minor. So even if we normalize the cross-sections to the NLO+NLL values, the change will be insignificant. Furthermore, the NLL-fast package used to evaluate NLO+NLL cross-sections is only available for 7, 8, 13, 14, 33 and 100 TeV energies and so for the sake of comparison between HL-LHC and HE-LHC we opted for LO calculation. The subsequent decay branching ratios of the gauginos are given in Table~\ref{tab5} which, along with the decay widths, are calculated by \code{SDECAY} and \code{HDECAY} operating within \code{SUSY-HIT}~\cite{Djouadi:2006bz}.

\section{Discovery channels for benchmarks \label{sec3}}

The SUSY signal  involving the direct production of $\tilde{\chi}_2^{0}\tilde{\chi}_1^{\pm}$ are simulated at LO using \code{MADGRAPH 2.6.0}~\cite{Alwall:2014hca} with the NNPDF23LO PDF set. The obtained parton-level sample is then passed to \code{PYTHIA8}~\cite{Sjostrand:2014zea} for showering and hadronization. Because we are interested in soft final states, no hard jets were added at the generator-level and so no matching/merging scheme is involved here. The necessary soft jets are added at the showering level. To give a boost to the final state particles, the initial (ISR) and the final state radiation (FSR) are relied upon for this purpose. The simulation of the direct production of a gluino pair is also carried out by \code{MADGRAPH} with up to one extra parton at the generator level. A five-flavor MLM matching is performed with the shower-kt scheme using \code{PYTHIA8} for showering and hadronization with the merging scale set at 120 GeV. Finally, ATLAS detector simulation and event reconstruction is performed by \code{DELPHES 3.4.1}~\cite{deFavereau:2013fsa} where clustering into jets is done by \code{FastJet}~\cite{Cacciari:2011ma} with a jet radius parameter 0.6 and using the anti-$k_T$ algorithm~\cite{Cacciari:2008gp}. \\        
For the 14 TeV backgrounds, we use the ones generated by the SNOWMASS group~\cite{Avetisyan:2013onh}. As for the 28 TeV samples, they are simulated at LO using \code{MADGRAPH 2.6.0} with the NNPDF30LO PDF set~\cite{Buckley:2014ana}. The cross-section is then multiplied by the appropriate K-factor so that it is close to its 
next-to-leading order (NLO) value. The resulting hard process is then passed on to \code{PYTHIA8} for showering and hadronization. To avoid double counting of jets, a five-flavor MLM matching is performed on the samples and the ATLAS detector simulation and event reconstruction is carried out by \code{DELPHES 3.4.1}. The SM backgrounds are classified as dominant and subdominant, where the subdominant backgrounds were given a K-factor of 1. \\
A large set of search analyses were performed on the generated events for each benchmark point. The analyses used \code{ROOT 6.08.06}~\cite{Antcheva:2011zz} to implement the constraints of the
search region for the signal regions involving leptons, jets and missing transverse energy in the final state. Since detector simulation is based on the ATLAS detector, most of the trigger level cuts used in this analysis are similar to those used by ATLAS, except for the trigger on the missing transverse energy. The leading and sub-leading jets are required to have $p_T > 20$ GeV and are reconstructed in $|\eta|<4.9$. Electrons and muons are required to have $p_T>10$ GeV where the former is reconstructed in the electromagnetic calorimeter with $|\eta|<2.47$ and the latter in the muon spectrometer with $|\eta|<2.5$. The trigger cut on the missing transverse energy used by ATLAS ranges from 100 to 150 GeV. However, in our analysis we lower the trigger on $E^{\rm miss}_T$ in order to capture soft final states, such that $E^{\rm miss}_T>70$ GeV.

\subsection{The single lepton channel \label{sec3.1}}

The analysis  of signatures  with leptons
has the advantage of being clean, i.e. contamination from QCD multi-jets is negligible. However, the downside of it is that the branching ratios for 
lepton signatures are relatively small.
The first channel we consider involves a single prompt light lepton in the final state, along with at least two jets and missing transverse energy. The standard model backgrounds pertaining to this final state include $W/Z$ + jets, $t\bar{t}$, diboson, $t$ + jets and $t+W/Z$. 
For both production processes, the single lepton comes mainly from the decay of a chargino. However, for the $\tilde{\chi}^0_2\tilde{\chi}^{\pm}_1$ pair production (benchmark points (g)-(j)) of Table~\ref{tab2},
 the final states are soft due to the small mass gap between the LSP and the chargino while for the gluino pair production case (benchmark points (a)-(f))
of Table~\ref{tab2},
 the electroweak gauginos are heavy ($\mathcal{O}(1)$ TeV) and hence we expect harder final states. For this reason, each signal region has two sets of selection criteria, one which targets soft final states (arising from a compressed spectrum)
 and given the suffix ``comp" and another targeting harder final states (arising from $\tilde g$ production)
 and given the suffix ``$\tilde{g}$". Note that the terms ``electroweakino production" and ``$\tilde\chi_2^0\tilde\chi_1^\pm$ production" are often used interchangeably in the text and thus refer to the same process.
A pre-selection cut on the missing transverse energy, $E^{\rm miss}_T>70$ GeV, is applied to both signal and background samples. Isolated leptons and jets are required to have $p_T>20$ GeV. The kinematic variables used for this signal region are:
\begin{itemize}
\item The lepton transverse mass
\begin{equation}
    m_{\rm T}^{\ell} =
    \sqrt{2p_{\rm T}^{\ell}E^{\rm miss}_{T}(1-\cos\Delta\phi(\ell,E^{\rm miss}_{T}))},
\end{equation}
which is used to reduce $t\bar{t}$ and $W+\rm jets$ backgrounds, where the $W$ boson decays leptonically, due to the fact that $m_T$ has a kinematical endpoint at the $W$ boson mass. In terms of jets, we could also define $m^{\rm min}_{T}(j_{1-2},E^{\rm miss}_T)$, the minimum of the transverse masses for the first two leading jets, which has the same effect on the above mentioned backgrounds.
\item The ratio $R$ defined as
\begin{equation}
R = \frac{E^{\rm miss}_{T}}{E^{\rm miss}_{T}+p_T^{\ell}}.
\end{equation}
Unlike the background, the signal tends to have a value of $R$ closer to one because of  the missing energy arising from the LSPs in the final state.
This ratio turns out to be  a powerful variable in discriminating signal from background.
\item $H_T$ is defined as the scalar sum of all the jets' transverse momenta in an event.
\item The effective mass, $m_{\rm eff}$, is given by
\begin{equation}
m_{\rm eff} = H_T + E^{\rm miss}_{T}+p_T^{\ell}.
\end{equation} 
Both $H_T$ and $m_{\rm eff}$ tend to have values higher than the background especially in processes involving gluino pair production.
\item The variable $E^{\rm miss}_T/\sqrt{H_T}$ is effective in eliminating possible multi-jet background events.
\item The Fox-Wolfram moments are given by~\cite{Fox:1978vu}
\begin{equation}
H_{\ell}=\sum_{ij}\frac{|\vec{p_i}||\vec{p_j}|}{E^2_{\rm vis}}P_{\ell}(\cos\theta_{ij}),
\end{equation}
where $\theta_{ij}$ is the separation angle between the two jets, $E_{\rm vis}$ is the total jet visible energy in an event and $P_{\ell}(x)$ are the Legendre polynomials. In particular, we use the normalized second Fox-Wolfram moment $H_{20}$ defined as $H_2/H_0$. This event shape observable is mostly effective for hard jets which is why it is only applied to the second set targeting gluino pair production. 
\end{itemize} 
The selection criteria for this SR using the above kinematic variables are listed in Table~\ref{tab6}. Each SR has three subsets labeled as SR-A, SR-B and SR-C which correspond to a variation of the ratio $R$. Observables and cuts that do not apply to a particular SR are left blank in the table. By comparing, for instance, cuts on $H_T$ and $m_{\rm eff}$, one can see that those observables have smaller values for the SR $1\ell$-comp which looks for soft final states, in comparison with $1\ell$-$\tilde{g}$ where much larger values are considered due to harder final states. The selection criteria is applied to the signal samples and to the 14 TeV and 28 TeV standard model backgrounds. The surviving events are used to determine the integrated luminosity for a $5\sigma$ $S/\sqrt{B}$ discovery. The results obtained for 14 TeV and 28 TeV are shown in Table~\ref{tab9} for all benchmark points of Table~\ref{tab2}. One can see that for LHC-14 most of the points cannot be discovered even with the HL-LHC as the required integrated luminosity exceeds 3000$\ifb$. Only points (a) and (g) are discoverable but require integrated luminosity greater than 1500$\ifb$.
 As for LHC-28, it is clear that all points can be discovered using the one lepton channel with an integrated luminosity as low as 32$\ifb$ (in SR-C for point (a)) which, given that such a machine may collect data at a rate of $\sim 820 \ifb$/year, may be attained within the first few weeks of operation. \\
For the gluino pair production case, the largest $\sigma_{\rm LO}\times \text{Br}(\tilde{\chi}^{\pm}_1\rightarrow\tilde{\chi}^0_1\ell^{\pm})$ is for point (a) followed by point (e), but those two points perform very differently with point (e) requiring much more integrated luminosity for discovery than the other points (except (f)). The reason is that for point (e), the chargino is almost degenerate with the LSP and hence the produced leptons are extremely soft and will not be reconstructed. In this case the single lepton is coming from the semi-leptonic decay of a top quark originating from the decay of a gluino (see Table~\ref{tab5}) and this has a small branching ratio. In figures~\ref{fig2}-\ref{fig4}, we exhibit the distributions in different kinematic variables for points (b), (e) and (f) at 1500$\ifb$ of integrated luminosity. The distributions show that even after applying the cuts, the signal is still buried under the 14 TeV backgrounds, while an excess is observed in all variables for the 28 TeV case. This is reflected in the calculated integrated luminosities of Table~\ref{tab9}. \\
Given that the obtained integrated luminosities for the 14 TeV case were larger than 3000$\ifb$ for most of the points, one must ask if optimizing the cuts for only the 14 TeV will improve the results. One could actually see that by looking at Fig.~\ref{fig5} which exhibits the distributions in $m_{\rm eff}$ and $R$ for benchmark point (g). Unlike the other points in Figs.~\ref{fig2}-\ref{fig4} where the signal is under the background, one can see that in Fig.~\ref{fig5} an excess is observed over the SM background. Those distributions have been plotted after all cuts, except $m_{\rm eff}$ and $R$, were applied. For the purpose of optimizing the cuts, we consider the SR $1\ell$-comp which applies to the points (g)-(j) of the $\tilde{\chi}^0_2\tilde{\chi}^{\pm}_1$ pair production. Instead of varying the cuts on $R$, we fix it to $R>0.85$ and vary the cuts on $m_{\rm eff}$ instead, which now become $m_{\rm eff}\in$ [200,450] for SR-A, [250,450] for SR-B and [250,420] for SR-C. The results obtained are tabulated in Table~\ref{tab1}.

\begin{table}[H]
	\centering
	\begin{tabulary}{\linewidth}{l|cccccccc}
    \hline\hline
    & \multicolumn{6}{c}{$\mathcal{L} ~(\times 10^3)$ for $5\sigma$ discovery at 14 TeV}  \\
    \hline
    & \multicolumn{3}{c}{SR $1\ell$ + jets} & \multicolumn{3}{c}{SR $1\ell$-Opt + jets} \\
	\hline
	Model & SR-A & SR-B & SR-C & SR-A & SR-B & SR-C \\
	\hline
  (g)  & 3.57 & 3.46 & 2.31 & 1.48 & 1.73 & 1.94 \\
  (h)  & 4.97 & 4.89 & 4.13 & 2.57 & 2.87 & 3.21 \\  
  (i)  & 5.23 & 5.21 & 4.79 & 3.13 & 3.67 & 4.13 \\
  (j)  & 6.16 & 6.09 & 6.04 & 3.48 & 3.93 & 4.39 \\
	\hline
	\end{tabulary}
	\caption{Comparison between the estimated integrated luminosity, in $(\times 10^3)$ fb$^{-1}$, for a 5$\sigma$ discovery at 14 TeV obtained before and after optimizing cuts in the single lepton channel.}
\label{tab1}
\end{table}

It is clear that there is a noticeable improvement in the integrated luminosities whereby points (g) and (h) are now discoverable at the HL-LHC. However, the integrated luminosities for those points at 28 TeV are still smaller. We note that no significant improvement is seen when trying to optimize the cuts for the 14 TeV case for the gluino production, i.e. points (a)-(f). From the analysis of~\cite{Aaboud:2018jiw} (Fig. 8d), it might appear that the parameter points (h)-(j) may be observable/excluded at the HL-LHC or even earlier. However, those points belong to the compressed spectrum where the chargino and LSP mass gap is small. For ATLAS and CMS experiments, this region is fairly challenging due to soft final states which requires much lower triggers than the ones already in use. In our work, the electroweakinos spectrum is based on a high scale model and thus the masses are not free parameters as assumed in simplified models used by ATLAS and CMS. Further, in those models, decay branching ratios are considered to be unity which is clearly not the case here as one can see from Table~\ref{tab5} where leptonic channels have smaller branching ratios. This implies a smaller cross-section and thus a larger integrated luminosity for discovery.
  
\subsection{The Two lepton channel \label{sec3.2}}

The second signal region (SR) consists of  two same flavor opposite sign (SFOS) leptons which originate from the decay of the second neutralino through an off-shell $Z$ boson (except for point (f) where the $Z$ boson is produced on-shell). The chargino will decay into jets through a $W$ boson (mostly off-shell) and an LSP (see Table~\ref{tab5}). Hence this signal region under consideration here (named $2\ell$-SFOS) consists of two SFOS light leptons, at least two jets and missing transverse energy. The leptonic decay of the second neutralino has a small branching ratio making this SR challenging and thus we do not expect it to perform as well as the single lepton channel. Further, the topology of this SUSY decay is very similar to standard model processes, i.e., the signal has the same shape as the SM background for some key kinematic variables which we discuss later.\\

The dominant standard model backgrounds for the SFOS
 final state consists of $t\bar{t}$, $Z/\gamma$ + jets, diboson and dilepton production from off-shell vector bosons. Subdominant backgrounds consists of Higgs production via gluon fusion ($ggF$ H) and vector boson fusion (VBF).
The selection criteria for this SR are presented in Table~\ref{tab7} where, again, two classes are considered: $2\ell$-SFOS-comp targeting soft final states from the $\tilde{\chi}^0_2\tilde{\chi}^{\pm}_1$ pair production (points (g)-(j)) and $2\ell$-SFOS-$\tilde{g}$ which is more suited for heavier electroweakinos resulting from the gluino decay (points (a)-(f)). Some of the discriminating variables used here overlap with the ones in the single lepton channel. The distinct ones include:
\begin{itemize}
\item The ratio $\mathcal{A}$ describes the $p_T$ asymmetry between the two leading jets and is given by
\begin{equation}
\mathcal{A}=\frac{p_T(j_1)-p_T(j_2)}{p_T(j_1)+p_T(j_2)}.
\end{equation}
This quantity is most effective when the mass gap between the NLSP and the LSP is small and thus will be used in this SR $2\ell$-SFOS-comp. 
\item The definition of the effective mass, $m_{\rm eff}$, for this SR region is modified to become
\begin{equation}
m_{\rm eff} = H_T + E^{\rm miss}_{T}+p_T^{\ell_1}+p_T^{\ell_2}.
\end{equation} 
\item The variable $M_{\rm T2}$~\cite{Lester:1999tx, Barr:2003rg, Lester:2014yga}  defined as    
\begin{equation}
    M_{\rm T2}=\min\left[\max\left(m_{\rm T}(\mathbf{p}_{\rm T1},\mathbf{q}_{\rm T}),
    m_{\rm T}(\mathbf{p}_{\rm T2},\,\mathbf{p}_{\rm T}^{\text{miss}}-
    \mathbf{q}_{\rm T})\right)\right],
    \label{mt2}
\end{equation}
where $\mathbf{q}_{\rm T}$ is an arbitrary vector chosen to find the appropriate minimum and $m_{\rm T}$ is the transverse mass given by 
\begin{equation}
    m_{\rm T}(\mathbf{p}_{\rm T1},\mathbf{p}_{\rm T2})=
    \sqrt{2(p_{\rm T1}\,p_{\rm T2}-\mathbf{p}_{\rm T1}\cdot\mathbf{p}_{\rm T2})}.
\end{equation}
This variable defined for the dilepton case is effective in reducing SM $t\bar{t}$ and $WW$ backgrounds.
\item The ratio $R$ for this SR becomes
\begin{equation}
R = \frac{E^{\rm miss}_{T}}{E^{\rm miss}_{T}+p_T^{\ell_1}+p_T^{\ell_2}}.
\end{equation}
\end{itemize}
In Table~\ref{tab7} each
 of the two signal regions is divided into three sub-regions, SR-A, SR-B and SR-C which correspond to variations in the separation between the SFOS leptons, $\Delta R_{\ell\ell}$, for $2\ell$-SFOS-comp and the lepton transverse mass, $m^{\ell}_T$, for $2\ell$-SFOS-$\tilde{g}$. Again, one can see   in Table~\ref{tab7} that
 harder cuts  are applied to the $2\ell$-SFOS-$\tilde{g}$ signal region. The cut on the dilepton invariant mass, $m_{\ell\ell}$ is kept low for $2\ell$-SFOS-comp which removes backgrounds coming from the decay of a $Z$ boson. A similar cut is applied to the SR $2\ell$-SFOS-$\tilde{g}$, named $Z$-veto, such that any dilepton mass within 10 GeV of the $Z$ boson pole mass is rejected. An additional veto on b-tagged and $\tau$-tagged jets is applied to the SR 2$\ell$-SFOS-comp.  After applying the cuts of Table~\ref{tab7}, the integrated luminosities required for a $5\sigma$ discovery are calculated and displayed in Table~\ref{tab10}. As expected, this SR does not perform as well as the single lepton channel did where much lower  integrated luminosities are observed.
Since the integrated luminosity scales like $E_{\rm CM}^2$, then doubling the center-of-mass energy means that the target integrated luminosity for the HE-LHC is around four times that of the HL-LHC. Hence, we expect that a total of up to 10 or 12$\iab$ of data to be collected at the HE-LHC. In Table~\ref{tab10}, we can see that points (g)-(j) can be discovered with an integrated luminosity of 267$\ifb$ for point (g) and $\sim 6650 \ifb$ for point (h). The rest of the points belonging to the gluino pair production case do not perform as well as in the single lepton channel. Points (e) and (f) are eliminated as they require more than 12$\iab$ for discovery in this particular SR. \\

In order to showcase the distribution of the signal versus the background for this SR, we display in Fig.~\ref{fig6} scatter plots for some key kinematic variables. In the top left panel, a scatter plot in the $(E^{\rm miss}_T+\sum p^{\ell}_T)$-$E^{\rm miss}_T$ plane is shown for point (d) at $\sqrt{s}=28$ TeV. One can see, as expected, that the signal (colored in orange) is clustered close to $R=1$ while the background is spread away which explains why a cut on $R$ close to 1 is needed to extract the signal. The top right panel shows a scatter plot in the $m_{\rm eff}$-$H_T$ plane for the same points after applying all the cuts except for the cuts on $m_{\rm eff}$ and $H_T$. The vertical and horizontal lines (with the arrows) show the position of the cuts on those variables. As a result, almost all of the signal points are maintained (in orange) and $t\bar{t}$ background (in red) is eliminated which has the largest remaining contribution after the cuts.          

\subsection{The three lepton channel \label{sec3.3}}

The final signal region to be considered is that of the three lepton channel. We require that two of the three selected leptons form a SFOS pair and no restrictions on the flavor and charge of the third lepton. For the SUSY signal, the SFOS pair comes from the decay of the second neutralino and the third lepton from the chargino decay. This channel also has a branching ratio smaller than the single lepton case. In cases where multiple SFOS pairs are present in an event, the transverse mass, $m_T$ is calculated for the unpaired lepton for each SFOS pairing and the minimum of the transverse masses, $m^{\rm min}_T$, is chosen and assigned to the $W$ boson mass. This variable and others used in this SR are listed in Table~\ref{tab8}. The variable $p^{\ell\ell\ell}_T$ is the transverse momentum of the three-lepton system, $M_{T2}$ is as defined by Eq.~(\ref{mt2}) and the ratio $R$ becomes
\begin{equation}
R = \frac{E^{\rm miss}_{T}}{E^{\rm miss}_{T}+\sum_{i=1}^3 p_T^{\ell_i}}
\end{equation}
and 
\begin{equation}
m_{\rm eff} = E^{\rm miss}_{T}+H_T+\sum_{i=1}^3 p_T^{\ell_i}.
\end{equation}
As before, a pre-selection cut, $E^{\rm miss}_T>70$ GeV, is applied. The three variations, SR-A, SR-B and SR-C, correspond to different values of the ratio $R$. 
The dominant backgrounds for this channel are SM $WZ$ diboson processes. Other contributing processes include trivector, $VVV$ ($V\in\{W,Z,\gamma\}$), $Z$ + jets, $t\bar{t}$, $t+W/Z$, dilepton production from off-shell vector bosons and Higgs production processes. 
After applying the cuts, the resulting integrated luminosities required for a $5\sigma$ discovery are calculated and shown in Table~\ref{tab11}. 
 The parameter points in Table~\ref{tab11} are discoverable  with luminosities ranging from $875\ifb$ for point (d) to  $5077\ifb$ for point (i)
 at 28 TeV while  none of these points would be  visible at 14 TeV. The other points are removed from the table since they require a minimum integrated luminosity of more than 3000$\ifb$ for 14 TeV and 12$\iab$ for 28 TeV. Despite having a higher production cross-section for a gluino pair, point (c) has a lower effective cross-section into the three-lepton final state, i.e. $\sigma_{(d)}(pp\rightarrow \tilde{g}\tilde{g})\times \text{Br}(\tilde{\chi}^0_2,\tilde{\chi}^{\pm}_0\rightarrow 3\ell)>\sigma_{(c)}(pp\rightarrow \tilde{g}\tilde{g})\times \text{Br}(\tilde{\chi}^0_2,\tilde{\chi}^{\pm}_0\rightarrow 3\ell) $ which explains why the integrated luminosity for point (d) is lower. 
Similar to the two lepton channel, we show a scatter plot of point (d) in the $(E^{\rm miss}_T+\sum p^{\ell}_T)$-$E^{\rm miss}_T$ plane for this SR. This is displayed in the bottom panel of Fig.~\ref{fig6} and again shows the signal  clustered near $R=1$.   

\section{Estimate of uncertainties \label{sec-uncern}}
In this section we give a rough estimate of the uncertainties associated with cross-section calculations which need to be propagated into the evaluated integrated luminosities. Theoretical systematic uncertainties arise from the renormalization and factorization scale variation, emission scale variation (for MLM merging), central scheme variation and parton distribution functions (PDF). For the gluino production cross-sections and the standard model backgrounds, we estimate $\sim 12\%$ and for the neutralino-chargino case $\sim 5\%$ systematics. Monte-Carlo simulation adds an uncertainty of $\sim 5\%$ for signal and $\sim 10\%$ for background. Experimental uncertainties are numerous but the largest contributions usually come from jet energy scale (JES) and jet energy resolution (JER), along with diboson production. Those uncertainties affect each SR differently. Based on~\cite{Aad:2015iea,Aad:2015jqa,Aaboud:2017leg,Aaboud:2017vwy,Aaboud:2017bac,Aaboud:2018jiw}, for SR-$1\ell$, experimental uncertainties can amount to $\sim 9\%$, for SR-$2\ell$-SFOS $\sim 12\%$ and SR-$3\ell$ $\sim 16\%$. The combined theoretical and experimental systematics for the signal and background are summarized in Table~\ref{tab12}. The range of values given for the signal systematics correspond to the neutralino-chargino production (lower bound) and gluino pair production (upper bound). The propagated uncertainties in the integrated luminosities are displayed in Table~\ref{tab13} for the leading and sub-leading SRs of the benchmark points of Table~\ref{tab2}.   

The analysis of this section shows that the dominant discovery channel for the class of models of Table~\ref{tab2} is most often the channel with a single lepton plus
jets. This is exhibited in  Table~\ref{tab13} where the discovery channel for all the parameter points is the single lepton plus jets channel,
with the exception of point (g) where the discovery channel is the two lepton plus jets channel, SR-2$\ell$ SFOS-A.  From Tables~\ref{tab1}, \ref{tab9}, \ref{tab10} and~\ref{tab11} only
the parameter points (a),  (g), (h) and (i) are discoverable at HL-LHC while all the parameter points (a)-(j) are discoverable at HE-LHC. For those points which are discoverable both at HL-LHC and HE-LHC, i.e., the points (a),  (g), (h) and (i), the time scale for discovery at HE-LHC will be much shorter. Thus the discovery of points (a), (g), (h) and (i)
would require a run of HL-LHC for $\sim 5$ yr for (a) and (g), and $\sim 8$ yr for (h) and (i). The run period for discovery of these at HE-LHC will be $\sim 2$  weeks for (a), $\sim 4$ months for (g), $\sim 1$ yr for (h) and $\sim 1.5$ yr for (i) using the projection that HE-LHC will collect 820 fb$^{-1}$ of data per year~\cite{HE-LHC-1}. 
As discussed in the paragraph above, our calculation of uncertainties is based mainly on the PDF variations and on experimental uncertainties reported by ATLAS and CMS for the same kind of analyses. The statistical uncertainties for LO processes as given by \code{MADGRAPH} are low. However, more realistic analyses of those will be done by experiments at HL-LHC and HE-LHC which may result in larger uncertainties. Dedicated studies of statistical and systematic uncertainties should make their way into the CERN's HL-LHC and HE-LHC yellow report scheduled to appear at the end of 2018. Even with high uncertainties, the main conclusion of our work should not be dramatically affected.

\section{Dark matter  \label{sec4}}
 We examine the detectability of the  benchmark points of Table~\ref{tab2} through direct detection of the neutralino which with R-parity is the dark matter candidate in SUGRA models 
over most of the parameter space of models~\cite{Arnowitt:1992aq}.
 The gaugino-higgsino content of the LSP determines the extent of detectability by virtue of the spin-independent (SI) and spin-dependent (SD) proton-neutralino cross-section. The neutralino is a mixture of a bino, wino and higgsinos, thus $\tilde{\chi}_0=\alpha\lambda^0+\beta\lambda^3+\gamma\tilde{H}_1+\delta\tilde{H}_2$, where $\alpha$ represents the bino content, $\beta$ the wino and $\delta$ and $\gamma$ the higgsino content. For points (a)-(f) the LSP is wino-like with $|\alpha|\leq 0.29$, $|\beta|\leq 0.97$, $|\gamma|\leq 0.016$ and $|\delta|\leq 0.004$, which explains the small dark matter relic density (see Table~\ref{tab3}), while for points (g)-(j) the LSP is bino-like, with $|\alpha|\leq 0.99$, $|\beta|\leq 0.28$ and $|\gamma|$ and $|\delta|$ are negligible. It must be noted that point (i) is obtained in a non-universal Higgs scenario and thus has a higher higgsino content such that $\sqrt{\gamma^2+\delta^2}\sim 0.06$. This has a significant effect on the SI and SD cross-sections which are displayed in Table~\ref{tab14} for all the benchmark points. Also in this table, we show the SI cross sections scaled by $\mathcal{R}=(\Omega h^2_{\tilde{\chi}^0_1})/(\Omega h^2)_{\rm PLANCK}$, with $\Omega h^2_{\tilde{\chi}^0_1}$ being the thermal neutralino relic density calculated in Table~\ref{tab2} and $(\Omega h^2)_{\rm PLANCK}$ is the dark matter relic density upper limited reported by the PLANCK experiment which amounts to $0.1197\pm 0.0022$. The analysis shows that those cross-sections lie below the current limits of 
XENON IT and LUX-ZEPLIN~\cite{Schumann:2015wfa,Cushman:2013zza} and often close to or even below the 
 neutrino floor~\cite{Strigari:2009bq} which is the threshold for detectability.  However, some of the parameter space may be accessible to  LUX-ZEPLIN in the future.

\section{Conclusions\label{sec5}}
The discovery of the Higgs boson mass at $\sim 125$ GeV indicates the  scale of weak scale supersymmetry lying in the several TeV region.
A scale of this size is needed to generate a large loop correction that can boost the tree level mass of the Higgs boson which in supersymmetry
 lies below the $Z$-boson mass to its experimentally observed value. The high scale of weak scale supersymmetry makes the observation of 
 supersymmetry at colliders more difficult pointing to the need for colliders with energies even  higher than the current LHC energy.
 Presently we are  in the LHC2 phase 
  with $\sqrt s=13$ TeV center-of-mass energy. This phase is expected to last till the end of  the year 2018 and it is projected that the CMS and ATLAS
  detectors will each have an integrated luminosity of 150$\ifb$ by then. Thus at the end of 2018, LHC will shut down for an upgrade  to LHC3 
  which will have a center-of-mass energy of 14 TeV, and will have its run in the period 2021-2023. In this period it is expected to collect 300$\ifb$ of
  data. The final upgrade  LHC4 is a high luminosity upgrade also referred to as HL-LHC, which will occur over  the period 2023-2026 
  and thereafter it will make a run over a  ten year period and  each detector is expected to collect up to 3000$\ifb$ of data. 
  A number of possibilities for the next collider after the LHC are under discussion. For supersymmetry, a proton-proton collider is the most relevant machine and here
  possibilities include a 100 TeV machine. At CERN a 100 TeV hadron collider will requires a 100 km circumference ring compared to the current 
  26.7 km circumference of the LHC collider.  A second  possible 100 km collider called Super proton-proton Collider (SppC)  is being considered  in China. 
  Recently a new proposal, HE-LHC, which would be a 28 TeV LHC has been 
  discussed~\cite{Benedikt:2018ofy,Zimmermann:2018koi,HE-LHC-1,HE-LHC-2,cern-report}.
   The main advantage 
  of this possibility is that this 28 TeV machine does not require a new tunnel and the upgrade in energy can be realized by use of more powerful
 16 T magnets using FCC technology compared to the 8.3 T magnets currently used by the LHC.\\

   In this work we have examined the potential for the discovery of supersymmetry within  a class of high scale models
  at 28 TeV. In the analysis we also make a comparison of the discovery potential at HE-LHC with that of HL-LHC.  
  The set of benchmarks we consider are based on well-motivated SUGRA models with radiative breaking of the electroweak 
  symmetry. The models have scalar masses in the several TeV region and gaugino masses which are much lighter
 consistent with the Higgs boson mass constraint and the constraint of relic density for the lightest neutralino. 
 All the benchmarks considered lie in regions which are not excluded by the current LHC data. The satisfaction of the relic 
 density constraint requires chargino co-annihilation which implies that the mass gap between the NLSP and the LSP is much 
 smaller than the LSP mass  which makes the detection of supersymmetry challenging which is typically the case for 
 models with compressed spectra. In the analysis we utilized several signature channels which include single lepton, two lepton, and three lepton channels
 accompanied with jets. Two sets of model points were analyzed, those which are beyond the reach of HL-LHC and, for comparison, also parameter points which are  
 discoverable at HL-LHC. For model points which are also discoverable at HL-LHC, it is found that their discovery at HE-LHC would take a much shorter 
 time reducing the run period of 5-8 yr at HL-LHC to  a run period of  few weeks to $\sim 1.5$ yr at HE-LHC. Thus  HE-LHC  is a  powerful tool 
 for the discovery of supersymmetry and deserves serious consideration.
Finally we note that while  the analysis was in progress HE-LHC energy was revised to 27 TeV. We do not expect our conclusions 
to be significantly affected by this revision. For some related works see also~\cite{Fowlie:2014awa,Chakrabortty:2015ika,Yamaguchi:2016oqz}.

\textbf{Acknowledgments:}
We thank Darien Wood  for discussions related to HE-LHC and  Baris Altunkaynak for discussions regarding some technical aspects of the analysis.
The analysis presented here was done using the resources of the high-performance  Cluster353 at the Advanced Scientific Computing Initiative (ASCI) and the Discovery Cluster at Northeastern University.  This research was supported in part by the NSF Grant PHY-1620575.

\section{Tables} 

\begin{table}[H]
\begin{center}
\begin{tabulary}{0.85\textwidth}{l|CCCCCC}
\hline\hline\rule{0pt}{3ex}
Model & $m_0$ & $A_0$ & $m_1$ & $m_2$ & $m_3$ & $\tan\beta$ \\
\hline\rule{0pt}{3ex}  
\!\!(a)  & 13998 & 30376 & 2155 & 1249 & 556 & 28 \\
(b)  & 9528  & 22200 & 2281 & 1231 & 573 & 34 \\
(c)  & 9288 & 20898 & 2471 & 1411 & 620 & 40 \\
(d)  & 28175 & 62830 & 2634 & 1541 & 751 & 41 \\
(e)  & 20335 & 44737 & 2459 & 1133 & 550 & 24 \\
(f)  & 22648  & 50505 & 2700 & 1585 & 675 & 15 \\
(g) & 16520  & 37224 & 385 & 274 & 1685 & 16 \\
(h)  & 48647 & 106537 & 537 & 432 & 2583 & 26 \\
(i)  & 14266 & -28965 & 371 & 224 & 2984 & 20 \\
(j)  & 41106  & 108520 & 687 & 599 & 7454 & 42 \\
\hline
\end{tabulary}\end{center}
\caption{Input parameters for the benchmark points used in this analysis. All points are obtained in non-universal gaugino  models  except for point (i) which is in a non-universal gaugino and Higgs scenario with $m_{H_d}=m_0$ and $m_{H_u}=18097$. All masses are in GeV.}
\label{tab2}
\end{table}

\begin{table}[H]
\begin{center}
\begin{tabulary}{1.1\textwidth}{l|CCCCCCC}
\hline\hline\rule{0pt}{3ex}
Model  & $h^0$ [GeV] & $\mu$ [TeV] & $\tilde\chi_1^0$ [$\times 10^2$ GeV] & $\tilde\chi_1^\pm$ [$\times 10^2$ GeV] & $\tilde t$ [TeV] & $\tilde g$ [TeV] & $\Omega^{\rm th}_{\tilde{\chi}^0_1} h^2$ \\
\hline\rule{0pt}{3ex} 
\!\!(a) & 124 & 8.02 & 9.73 & 10.6 & 4.73 & 1.36 & 0.039 \\
(b) & 125 & 6.29 & 10.2 & 10.3 & 2.08 & 1.40 & 0.035 \\
(c) & 123 & 5.59 & 11.1 & 11.9 & 2.88 & 1.51 & 0.048 \\
(d) & 124 & 15.5 & 11.9 & 12.7 & 10.0 & 1.75 & 0.048 \\
(e) & 124 & 11.7 & 9.48 & 9.48 & 6.78 & 1.33 & 0.020 \\
(f) & 124 & 13.7 & 12.4 & 13.5 & 6.98 & 1.62 & 0.112 \\
(g) & 124 & 10.4 & 1.34 & 1.51 & 5.27 & 3.93 & 0.121 \\
(h) & 124 & 26.1 & 1.54 & 1.76 & 18.6 & 5.88 & 0.105  \\
(i) & 124 & 1.15 & 1.65 & 1.89 & 4.17 & 6.71 & 0.114 \\
(j) & 125 & 29.7 & 1.62 & 1.87 & 10.4 & 15.6 & 0.105 \\
\hline
\end{tabulary}\end{center}
\caption{The Higgs boson ($h^0$) mass, the $\mu$ parameter and 
some relevant sparticle masses, and the relic density for the benchmark points of Table~\ref{tab2}. }
\label{tab3}
\end{table}

\begin{table}[H]
\centering
\begin{tabulary}{2.0\textwidth}{l|CC}
\hline\hline\rule{0pt}{3ex}
 & $\sqrt{s} = 14$ TeV & $\sqrt{s} = 28$ TeV \\
\hline
Model & \multicolumn{2}{c}{$\sigma_{\rm LO}^{\rm matched}(pp\rightarrow\tilde{g}\tilde{g})$} \\
\hline
(a) & 0.029 & 0.67 \\
(b) & 0.023 & 0.55 \\
(c) & 0.012 & 0.34 \\
(d) & 0.004 & 0.14 \\
(e) & 0.036 & 0.79 \\
(f) & 0.007 & 0.23 \\
\hline
& \multicolumn{2}{c}{$\sigma_{\rm LO}(pp\rightarrow\tilde\chi_2^0\tilde\chi_1^\pm)$} \\
 \hline
(g) & 4.11 & 10.34 \\
(h) & 2.25 & 5.86 \\
(i) & 1.65 & 4.38 \\
(j) & 1.78 & 4.71  \\
\hline
\end{tabulary}
\caption{Top table: Production cross-section for  the two gluino production cross section $pp\to \tilde g \tilde g$ to leading order
at $\sqrt s=14$ TeV and $\sqrt s=28$ TeV center-of-mass energy  for the top six 
benchmarks points of Table~\ref{tab2} where the cross sections are in pico-barns.  Bottom table: Same as the top table except the production cross section 
is for the second neutralino and the light chargino process $ (pp\rightarrow\tilde\chi_2^0\tilde\chi_1^\pm)$ for the four bottom benchmark points of 
Table~\ref{tab2}.  }
\label{tab4}
\end{table}  

\begin{table}[H]
\begin{center}
\begin{tabulary}{2.00\textwidth}{l|CCCC}
\hline\hline\rule{0pt}{3ex}
Model & $\tilde{g}\to\tilde\chi_1^0 q\bar{q}$ & $\tilde{g}\to\tilde\chi_1^{\pm}q_i\bar{q}_j$  & $\tilde{g}\to\tilde\chi_2^0 q\bar{q}$ & $\tilde{\chi}_1^{\pm}\to\tilde\chi_1^0 W^{\pm}$\footnotemark \\
& \multicolumn{3}{c}{$q\in\{u,d,c,s,t,b\}$} &  \\
\hline
(a) & 0.66 & 0.29 & 0.05 & 1.0 \\
(b) & 0.33 & 0.53 & 0.13 & 0.32 \\
(c) & 0.63 & 0.33 & 0.04 & 0.33 \\
(d) & 0.48 & 0.40 & 0.12 & 1.0 \\
(e) & 0.36 & 0.63 & 0.01 & 0.25 \\
(f) & 0.73 & 0.20 & 0.07 & 1.0 \\
\hline
 & $\tilde\chi_2^0\to\tilde\chi_1^0 q\bar{q}$ & $\tilde\chi_2^0\to\tilde\chi_1^0\ell\bar{\ell}$  & $\tilde{\chi}_1^{\pm}\to\tilde\chi_1^0 q_i\bar{q_j}$ & $\tilde{\chi}_1^{\pm}\to\tilde\chi_1^0 \ell^{\pm}\nu_{\ell}$\\
& $q\in\{u,d,c,s\}$ & $\ell\in\{e,\mu,\tau,\nu\}$ & $q\in\{u,d,c,s,b\}$ & $\ell\in\{e,\mu,\tau\}$ \\
\hline\rule{0pt}{3ex}
\!\!(g) & 0.88 & 0.12 & 0.67 & 0.33 \\
(h) & 0.84 & 0.16 & 0.67 & 0.33 \\
(i) & 0.68 & 0.32 & 0.67 & 0.33 \\
(j) & 0.94 & 0.06 & 0.67 & 0.33 \\
\hline
\end{tabulary}
\end{center}
\caption{Branching ratios for the dominant decays of $\tilde\chi_2^0$ and $\tilde\chi_1^\pm$ for benchmark points of Table~\ref{tab2} where $q_i\bar{q}_j=\{(u\bar{d}), (c\bar{s}), (t\bar{b})\}.$}
\label{tab5}
\end{table}
\footnotetext{For Br($\tilde{\chi}_1^{\pm}\to\tilde\chi_1^0 W^{\pm})=1$, the $W$ boson is produced on-shell}

\begin{table}[H]
	\begin{center}
	\begin{tabulary}{\linewidth}{lccc|ccc}
    \hline\hline
	 & \multicolumn{3}{c}{SR 1$\ell$-comp} & \multicolumn{3}{c}{SR 1$\ell$-$\tilde{g}$} \\
    \cline{2-4} \cline{5-7} 
    Requirement & SR-A & SR-B & SR-C & SR-A & SR-B & SR-C \\
    \hline
    $N_{\rm jets} \geq $
    & 2 & 2 & 2 & 2 & 2 & 2 \\
    $E^{\text{miss}}_T\text{ (GeV)} > $
    &  &  &  & 150 & 150 & 150 \\
    $H_{T} ~\text{(GeV)} $
    & $< 250$ & $< 250$ & $< 250$ & $> 600$ & $> 600$ & $> 600$ \\
    $E^{\rm miss}_{T}/\sqrt{H_{T}} ~(\text{GeV}^{1/2})> $
    & 7 & 7 & 7 & 10 & 10 & 10 \\
    $m_{\rm eff}\text{ (GeV)}$
    & $< 350$ & $< 350$ & $< 350$ & $> 800$ & $> 800$ & $> 800$ \\
    $R > $
    & 0.6 & 0.7 & 0.85  & 0.7 & 0.8 & 0.85 \\
    $H_{20} < $
    &  &  &  & 0.5 & 0.5 & 0.5 \\
    $p_{T}(j_2) \text{ (GeV)}>$
    &  &  &  & 110 & 110 & 110 \\
    $p_{T}(j_3) \text{ (GeV)}>$
    &  &  &  & 80 & 80 & 80 \\
    $p_{T}(j_4) \text{ (GeV)}>$
    &  &  &  & 50 & 50 & 50 \\
    $m^{\ell}_T \text{ (GeV)}>$
    &  &  &  & 100 & 100 & 100 \\
    $m^{\rm min}_T (j_{1-2},E^{\rm miss}_{T})\text{ (GeV)}>$
    &  &  &  & 200 & 200 & 200 \\
	\hline
	\end{tabulary}\end{center}
	\caption{The selection criteria (SR) used for the single lepton + jets signal region. The SR 1$\ell$-comp targets soft final states resulting from the electroweakino ($\tilde\chi_2^0\tilde\chi_1^\pm$) production and 1$\ell$-$\tilde{g}$ targets final states from gluino production.}
	\label{tab6}
\end{table}

\begin{table}[H]
	\begin{center}
	\begin{tabulary}{\linewidth}{lccc|ccc}
    \hline\hline
	 & \multicolumn{3}{c}{2$\ell$-SFOS-comp} & \multicolumn{3}{c}{2$\ell$-SFOS-$\tilde{g}$} \\
    \cline{2-4} \cline{5-7}
    Requirement & SR-A & SR-B & SR-C & SR-A & SR-B & SR-C \\
    \hline
    $N_{\rm jets} \geq $
    & 2 & 2 & 2 & 2 & 2 & 2 \\
    $E^{\text{miss}}_T\text{ (GeV)}$
    & $<150$ & $<150$ & $<150$ & $>150$ & $>150$ & $>150$ \\
    $m^{\ell}_T \text{ (GeV)}$
    & $<80$ & $<80$ & $<80$ & $>150$ & $>180$ & $>200$ \\    
    $p_{T}(j_1) \text{ (GeV)}$
    & $<90$ & $<90$ & $<90$ & $>120$ & $>120$ & $>120$ \\
    $p_{T}(j_2) \text{ (GeV)}$
    & $<48$ & $<48$ & $<48$ & $>80$ & $>80$ & $>80$ \\
    $R > $
    & 0.7 & 0.7 & 0.7 & 0.8 & 0.8 & 0.8 \\
    $A < $
    & 0.4 & 0.4 & 0.4  \\
    $m_{\ell\ell}\text{ (GeV)}<$
    & 25 & 25 & 25 & \multicolumn{3}{c}{$Z$-veto} \\
    $H_{T} \text{(GeV)}$
    & $<190$ & $<190$ & $<190$ & $>500$ & $>500$ & $>500$ \\
    $m_{\rm eff}\text{ (GeV)}>$
    & 180 & 180 & 180 & 900 & 900 & 900 \\
    $m_{\rm eff}\text{ (GeV)}<$
    & 400 & 400 & 400 &  &  &  \\
    $\Delta R_{\ell\ell}\text{ (rad)}<$
    & 0.4 & 1.0 & 2.5 & & & \\
    $M_{T_2}^{\rm dijet} \text{ (GeV)} > $
    &  &  &  & 700 & 700 & 700 \\
    $M_{T_2}^{\rm dilepton} \text{ (GeV)} > $
    &  &  &  & 600 & 600 & 600 \\
	\hline
	\end{tabulary}\end{center}
	\caption{The selection criteria used for the two-lepton same flavor opposite sign signal region. The SR 2$\ell$-SFOS-comp targets soft final states resulting from the electroweakino production and 2$\ell$-SFOS-$\tilde{g}$ targets final states from gluino production.}
	\label{tab7}
\end{table}

\begin{table}[H]
	\begin{center}
	\begin{tabulary}{\linewidth}{lccc|ccc}
    \hline\hline
	 & \multicolumn{3}{c}{SR 3$\ell$-comp} & \multicolumn{3}{c}{SR 3$\ell$-$\tilde{g}$} \\
    \cline{2-4} \cline{5-7}
    Requirement & SR-A & SR-B & SR-C & SR-A & SR-B & SR-C \\
    \hline
    $E^{\text{miss}}_T\text{ (GeV)}>$
    &  &  &  & 150 & 150 & 150 \\
    $m^{\rm min}_T \text{ (GeV)}>$
    &  &  &  & 100 & 100 & 100 \\    
    $p_{T}^{\ell\ell\ell} \text{ (GeV)}<$
    & 60 & 60 & 60 & 150 & 150 & 150 \\
    $R > $
    & 0.45 & 0.5 & 0.55 & 0.55 & 0.6 & 0.65 \\
    $m_{\rm eff}\text{ (GeV)}$
    & $<500$ & $<500$ & $<500$ & $>650$ & $>650$ & $>650$ \\
    $M_{T_2}^{\rm dijet} \text{ (GeV)} > $
    & 200 & 200 & 200 &  &  &  \\
	\hline
	\end{tabulary}\end{center}
	\caption{The selection criteria used for the three-lepton signal region. The SR 3$\ell$-comp targets soft final states resulting from the electroweakino production and 3$\ell$-$\tilde{g}$ targets final states from gluino production. A $Z$-veto is applied to both SRs.}
	\label{tab8}
\end{table}

\begin{table}[H]
	\centering
	\begin{tabulary}{\linewidth}{l|cccccccc}
    \hline\hline
    & \multicolumn{6}{c}{$\mathcal{L}$ for $5\sigma$ discovery in 1$\ell$ + jets}  \\
    \hline
    & \multicolumn{3}{c}{$\mathcal{L} ~(\times 10^3)$ at 14 TeV} & \multicolumn{3}{c}{$\mathcal{L} ~(\times 10^3)$ at 28 TeV} \\
	\hline
	Model & SR-A & SR-B & SR-C & SR-A & SR-B & SR-C \\
	\hline
  (a)  & 2.30 & 1.84 & 1.60 & 0.050 & 0.037 & 0.032 \\
  (b)  & 3.44 & 3.26 & 3.45 & 0.054 & 0.048 & 0.047 \\
  (c)  & 9.84 & 8.32 & 7.58 & 0.124 & 0.096 & 0.082 \\
  (d)  & 16.0 & 14.2 & 13.0 & 0.129 & 0.105 & 0.096 \\
  (e)  & 8.25 & 8.64 & 9.56 & 0.228 & 0.206 & 0.220 \\
  (f)  & 43.6 & 38.0 & 40.1 & 0.443 & 0.358 & 0.352 \\
  (g)  & 3.57 & 3.46 & 2.31 & 0.798 & 0.708 & 0.562 \\
  (h)  & 4.97 & 4.88 & 4.13 & 1.08 & 0.983 & 0.996 \\  
  (i)  & 5.23 & 5.21 & 4.79 & 1.35 & 1.22 & 1.22 \\
  (j)  & 6.16 & 6.09 & 6.04 & 1.40 & 1.32 & 1.52 \\
	\hline
	\end{tabulary}
	\caption{Comparison between the estimated integrated luminosity for a 5$\sigma$ discovery at 14 TeV and 28 TeV for supersymmetry for the parameter {set}  of Table~\ref{tab2}, using the selection criteria of Table~\ref{tab6}, where the minimum integrated luminosity needed for a $5\sigma$ discovery is given in $(\times 10^3)$ fb$^{-1}$.}
\label{tab9}
\end{table}

\begin{table}[H]
	\centering
	\begin{tabulary}{\linewidth}{l|cccccccc}
    \hline\hline
    & \multicolumn{6}{c}{$\mathcal{L}$ for $5\sigma$ discovery in 2$\ell$ SFOS}  \\
    \hline
    & \multicolumn{3}{c}{$\mathcal{L} ~(\times 10^3)$ at 14 TeV} & \multicolumn{3}{c}{$\mathcal{L} ~(\times 10^3)$ at 28 TeV} \\
	\hline
	Model & SR-A & SR-B & SR-C & SR-A & SR-B & SR-C \\
	\hline
  (g)  & 4.33 & 2.87 & 5.56 & 0.267 & 0.309 & 0.708 \\
  (d)  & 248 & 265 & 263 & 1.14 & 1.63 & 1.82 \\
  (a)  & 73.8 & 71.9 & 78.6  & 1.27 & 1.88 & 2.34 \\
  (b)  & 64.2 & 59.7 & 80.9  & 2.27 & 2.16 & 3.39 \\
  (c)  & 148 & 165 & 242  & 2.32 & 2.56 & 2.86 \\
  (i)  & 4.43 & 3.10 & 2.52 & 2.69 & 2.16 & 2.32 \\
  (h)  & 10.6 & 5.69 & 6.04 & 6.65 & ... & 7.84 \\
  (j)  &  & 30.3 & 37.6 & 10.2 & 3.64 & 3.00 \\
	\hline
	\end{tabulary}
	\caption{Comparison between the estimated integrated luminosity for a 5$\sigma$ discovery at 14 TeV and 28 TeV for supersymmetry for the parameter {set}  of Table~\ref{tab2}, using the selection criteria of Table~\ref{tab7}, where the minimum integrated luminosity needed for $5\sigma$ discovery is given in $(\times 10^3)$ fb$^{-1}$. Points that are not discoverable, i.e., require a minimum integrated luminosity which exceeds 3000$\ifb$ for $\sqrt s=14$ TeV and 10 ab$^{-1}$ for $\sqrt s=28$ TeV are not exhibited. Blank entries mean that no events have passed the cuts.}
\label{tab10}
\end{table}

\begin{table}[H]
	\centering
	\begin{tabulary}{\linewidth}{l|cccccccc}
    \hline\hline
    & \multicolumn{6}{c}{$\mathcal{L}$ for $5\sigma$ discovery in 3$\ell$ channel}  \\
    \hline
    & \multicolumn{3}{c}{$\mathcal{L} ~(\times 10^3)$ at 14 TeV} & \multicolumn{3}{c}{$\mathcal{L} ~(\times 10^3)$ at 28 TeV} \\
	\hline
	Model & SR-A & SR-B & SR-C & SR-A & SR-B & SR-C \\
	\hline
  (d)  & ... & ... & ... & 1.12 & 0.875 & 0.108 \\
  (c)  & ... & ... & ... & 2.27 & 1.77 & 1.61 \\
  (g)  & ... & ... & ... & 5.94 & 5.35 & ... \\
  (i)  & ... & ... & ... & 6.63 & 5.97 & 5.08 \\
	\hline
	\end{tabulary}
	\caption{Comparison between the estimated integrated luminosity for a 5$\sigma$ discovery at 14 TeV and 28 TeV for supersymmetry for the parameter {set}  of Table~\ref{tab2}, using the selection criteria of Table~\ref{tab8}, where the minimum integrated luminosity needed for $5\sigma$ discovery is given in $(\times 10^3)$ fb$^{-1}$. Entries with ... correspond to integrated luminosities $\mathcal{O}(10^6)\ifb$ and thus not displayed.}
\label{tab11}
\end{table}

\begin{table}[H]
\begin{center}
\begin{tabulary}{0.85\textwidth}{lCC}
\hline\hline\rule{0pt}{3ex}
Signal region & Signal systematics & Background systematics \\
\hline\rule{0pt}{3ex}
\!\!SR-$1\ell$ & 11.4-15.8\% & 18.0\%  \\
SR-$2\ell$-SFOS  & 13.9-17.7\% & 19.7\%  \\
SR-$3\ell$ & 17.5-20.6\% & 22.4\%  \\
\hline
\end{tabulary}\end{center}
\caption{Total estimated systematic uncertainties on signal
and background for the three signal regions.}
\label{tab12}
\end{table}

\begin{table}[H]
\begin{center}
\begin{tabulary}{1.00\textwidth}{lCCCCC}
\hline\hline\rule{0pt}{3ex}
Model & Leading SR & $\mathcal{L}$ (fb$^{-1}$) & Sub-leading SR & $\mathcal{L}$ (fb$^{-1}$)\\
\hline\rule{0pt}{3ex}
\!\!(a) & SR-$1\ell$-C & $32\pm 5$ & SR-$1\ell$-B & $37\pm 6$ \\
(b) & SR-$1\ell$-C & $47\pm 7$ & SR-$1\ell$-B & $48\pm 8$ \\
(c)& SR-$1\ell$-C & $82\pm 10$ & SR-$1\ell$-B & $96\pm 13$ \\
(d)& SR-$1\ell$-C & $96\pm 13$ & SR-$1\ell$-B & $105\pm 15$ \\
(e)& SR-$1\ell$-B & $206\pm 22$ & SR-$1\ell$-C & $220\pm 21$ \\
(f)& SR-$1\ell$-C & $352\pm 37$ & SR-$1\ell$-B & $358\pm 34$ \\
(g)& SR-$2\ell$ SFOS-A & $267\pm 23$ & SR-$2\ell$ SFOS-B & $309\pm 32$ \\
(h)& SR-$1\ell$-B & $983\pm 96$ & SR-$1\ell$-C & $996\pm 81$ \\
(i)& SR-$1\ell$-C & $ (1.22\pm 0.15)\times 10^3$ & SR-$1\ell$-B & $(1.22\pm 0.14)\times 10^3$ \\
(j)& SR-$1\ell$-B & $(1.32\pm 0.17)\times 10^3$ & SR-$1\ell$-A & $(1.40\pm 0.19)\times 10^3$ \\
\hline
\end{tabulary}\end{center}
\caption{
The overall minimum integrated luminosities needed for $5\sigma$ discovery at the HE-LHC, displayed with their total uncertainty, using the
leading and the sub-leading signal regions for benchmarks of Table~\ref{tab2}, including all signal regions discussed.}
\label{tab13}
\end{table}

\begin{table}[H]
\begin{center}
\begin{tabulary}{0.85\textwidth}{lCCC}
\hline\hline\rule{0pt}{3ex}
Model & $\sigma^{\text{SI}}_{p,\chi^0_1}$ & $\sigma^{\text{SD}}_{p,\chi^0_1}$ & $\mathcal{R}\times\sigma^{\text{SI}}_{p,\chi^0_1}$\\
\hline\rule{0pt}{3ex}
\!\!(a)& $1.43\times 10^{-48}$ & $5.36\times 10^{-46}$ & $4.66\times 10^{-49}$ \\
(b)& $1.56\times 10^{-48}$ & $7.86\times 10^{-46}$ & $4.60\times 10^{-49}$ \\
(c)& $5.09\times 10^{-48}$ & $2.30\times 10^{-45}$ & $2.04\times 10^{-48}$ \\
(d)& $9.22\times 10^{-50}$ & $2.16\times 10^{-47}$ & $3.70\times 10^{-50}$ \\
(e)& $2.49\times 10^{-49}$ & $6.47\times 10^{-47}$ & $4.16\times 10^{-50}$ \\
(f)& $7.05\times 10^{-49}$ & $6.90\times 10^{-47}$ & $6.60\times 10^{-49}$ \\
(g)& $4.95\times 10^{-50}$ & $1.23\times 10^{-47}$ & $5.01\times 10^{-50}$ \\ 
(h)& $6.79\times 10^{-51}$ & $1.09\times 10^{-48}$ & $5.96\times 10^{-51}$ \\
(i)& $3.18\times 10^{-47}$ & $3.70\times 10^{-43}$ & $3.03\times 10^{-47}$ \\
(j)& $2.33\times 10^{-51}$ & $6.98\times 10^{-49}$ & $2.05\times 10^{-51}$ \\ 
\hline
\end{tabulary}
\caption{Proton--neutralino spin-independent ($\sigma^{\text{SI}}_{p,\chi^0_1}$ and $\mathcal{R}\times\sigma^{\text{SI}}_{p,\chi^0_1}$ with $\mathcal{R}=(\Omega h^2_{\tilde{\chi}^0_1})/(\Omega h^2)_{\rm PLANCK}$) and spin-dependent ($\sigma^{\text{SD}}_{p,\chi^0_1}$) cross-sections in units of cm$^{2}$ for the benchmarks  of Table~\ref{tab2}.}
\label{tab14}
\end{center}
\end{table}

\section{Figures}

\begin{figure}[H]
 \centering
 	\includegraphics[width=0.45\textwidth]{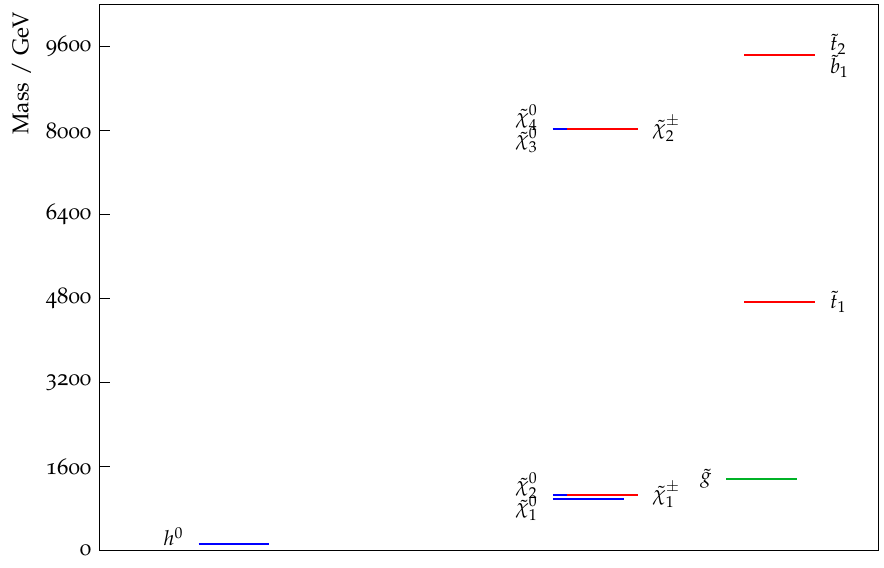}
   \includegraphics[width=0.45\textwidth]{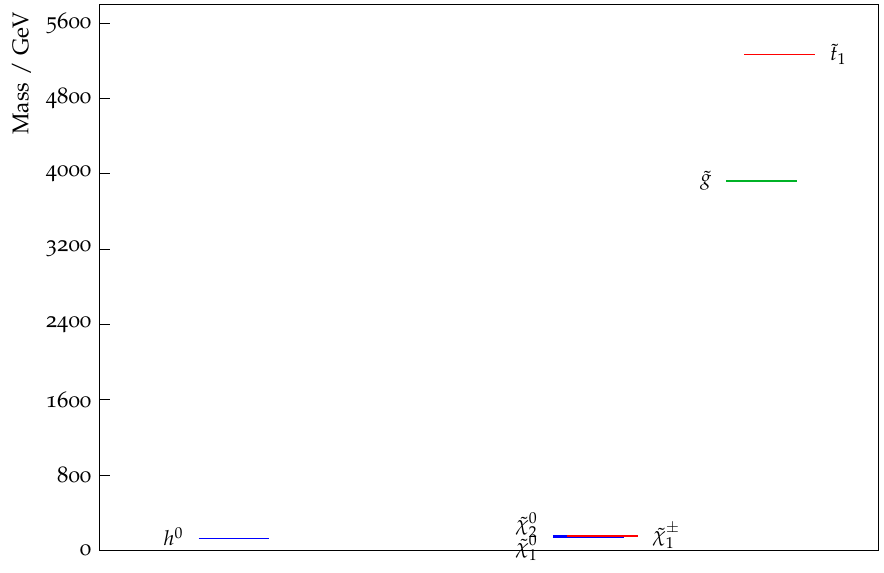}
    \caption{Left panel: The sparticle spectrum for the benchmark (a) of Table~\ref{tab2} with a light gluino.  Right panel: same but for point (g) with a heavier gluino.}
	\label{spectrum}
\end{figure}

\begin{figure}[H]
 \centering
   \includegraphics[width=0.45\textwidth]{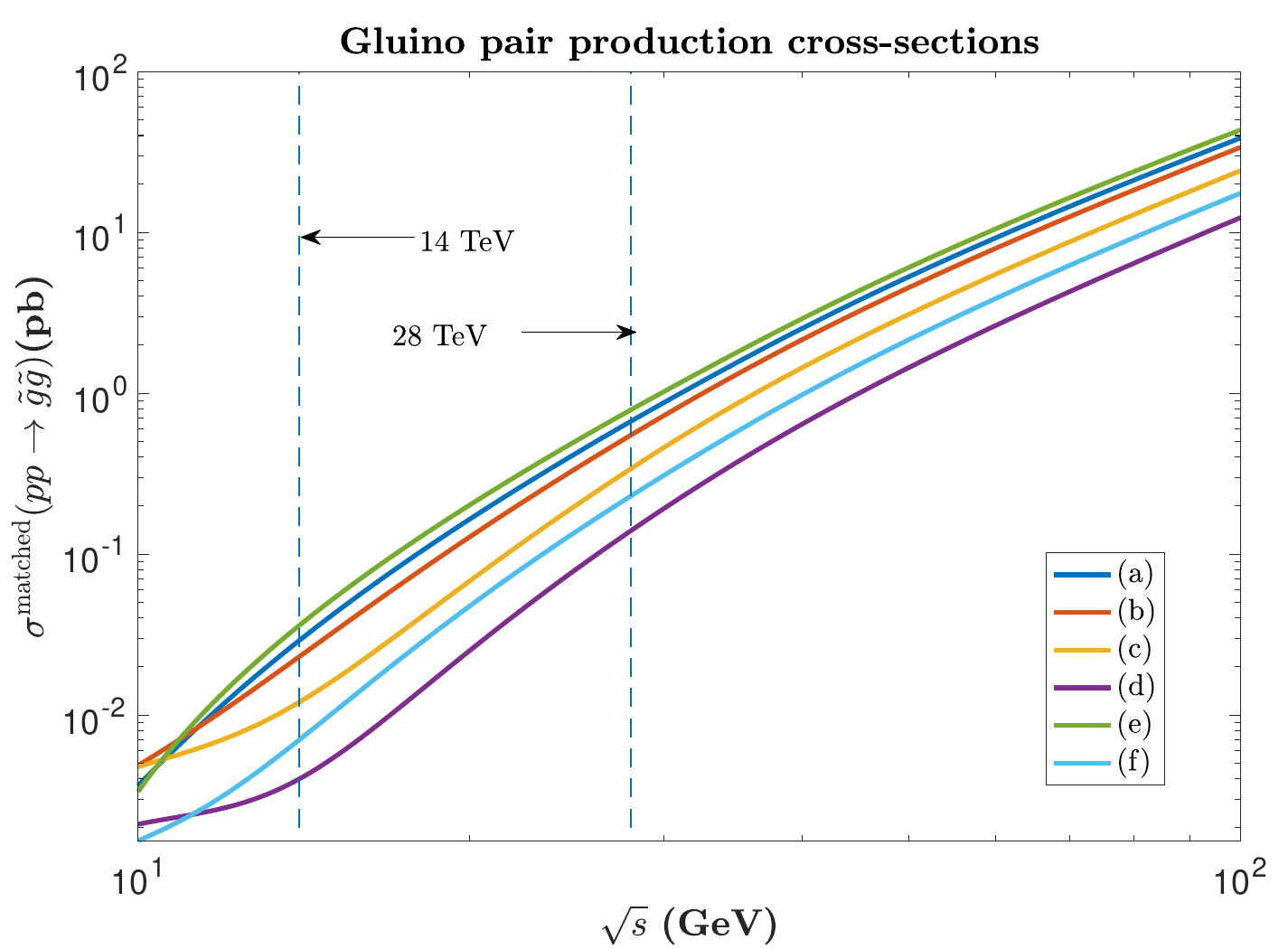} 
 	\includegraphics[width=0.45\textwidth]{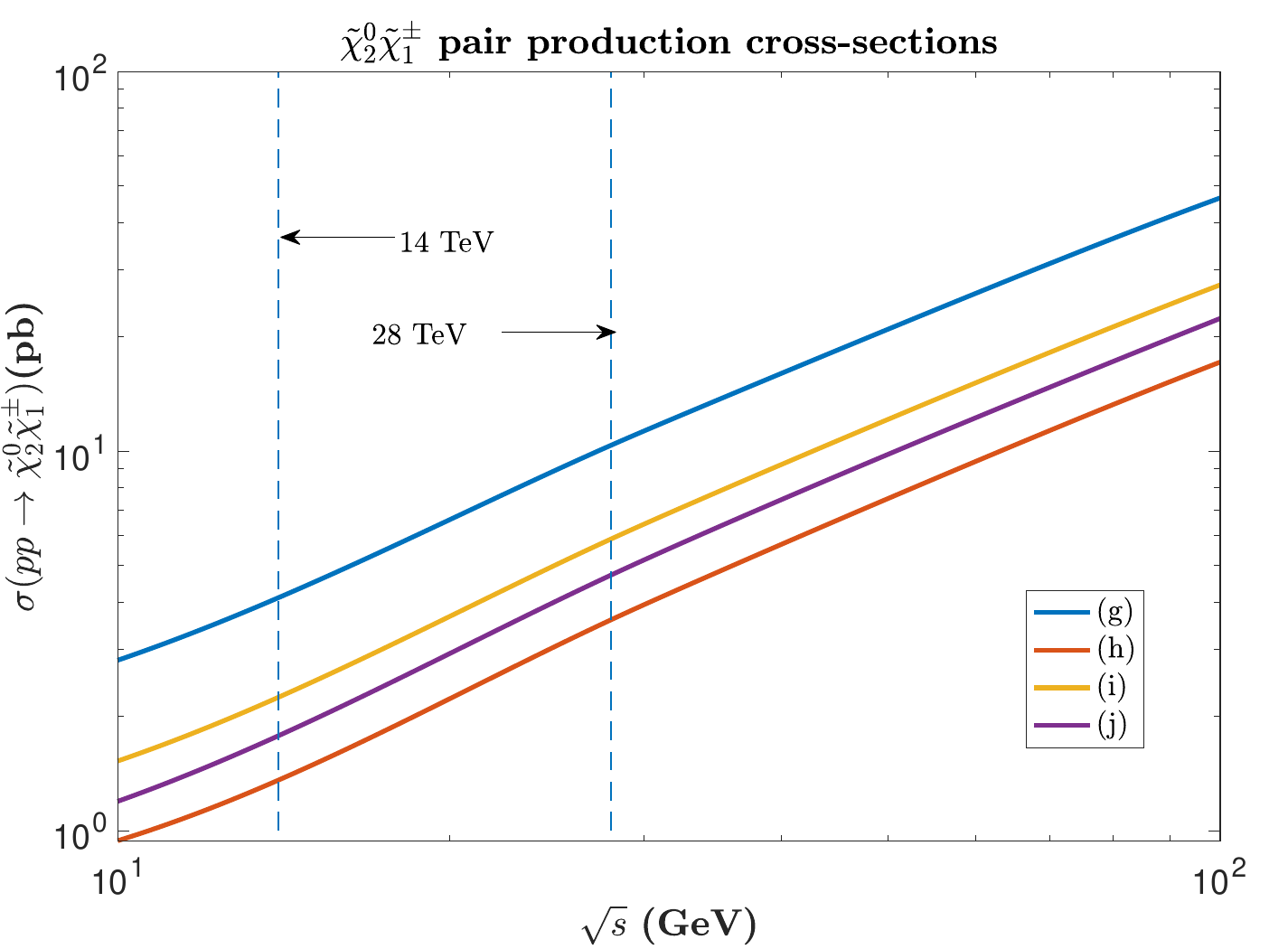}
   \caption{Left panel: $\tilde g \tilde g$ production cross section a function of the center-of-mass energy for $pp$ collisions for the
    parameter points (a), (b), (c), (d), (e), (f) of Table~\ref{tab2}. Right panel: $\tilde{\chi}^0_2 \tilde{\chi}^{\pm}$ production cross section a function of the center-of-mass energy for $pp$ collisions for the
    parameter points (g), (h), (i), (j) of Table~\ref{tab2}.}
	\label{fig1}
\end{figure}

\begin{figure}[H]
 \centering
 	\includegraphics[width=0.40\textwidth]{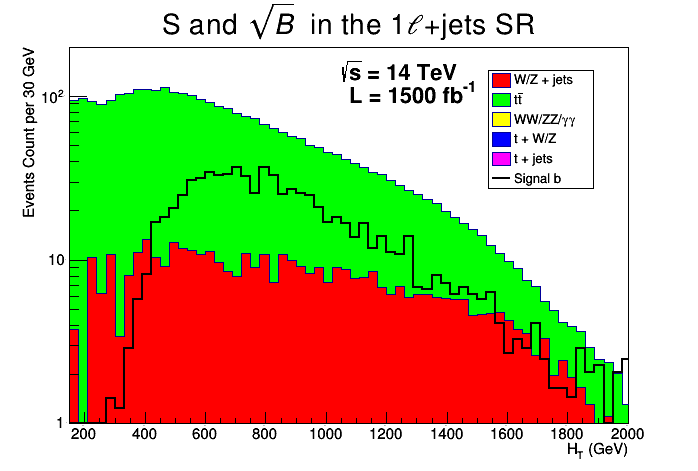}
   \includegraphics[width=0.40\textwidth]{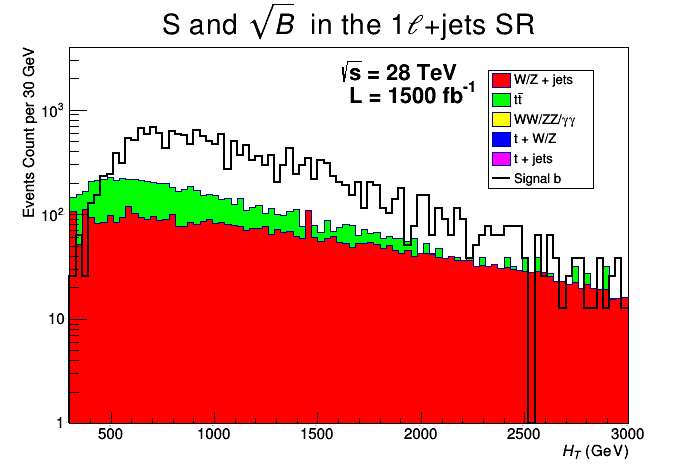}
   \includegraphics[width=0.40\textwidth]{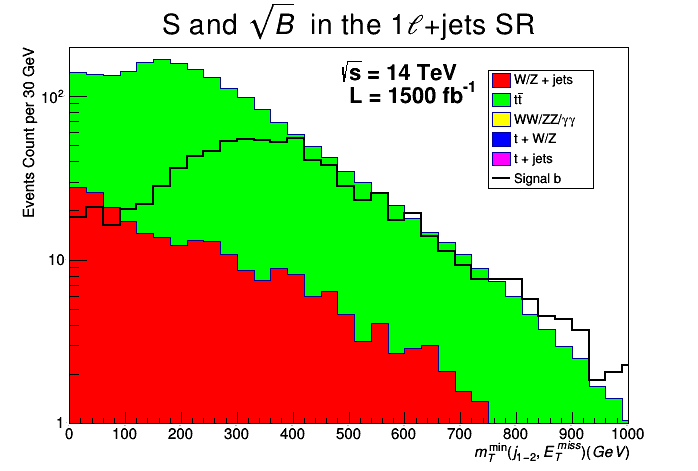}
   \includegraphics[width=0.40\textwidth]{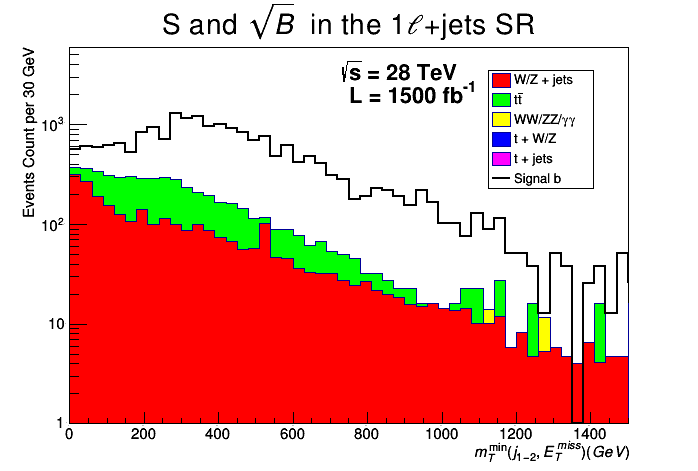}
    \caption{Distributions in the kinematic variable $H_T$ (top) and $m^{\rm min}_{T}(j_{1-2},E^{\rm miss}_{T})$ (bottom) for benchmark point (b) at 14 TeV (left) and 28 TeV (right) in the single lepton channel.}
	\label{fig2}
\end{figure}

\begin{figure}[H]
 \centering
 	\includegraphics[width=0.40\textwidth]{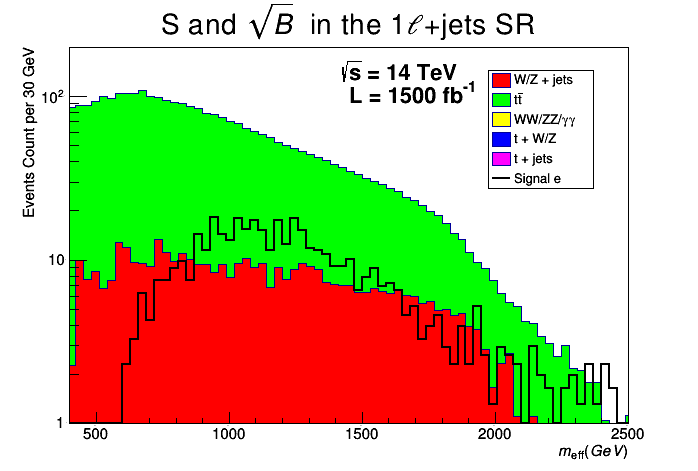}
   \includegraphics[width=0.40\textwidth]{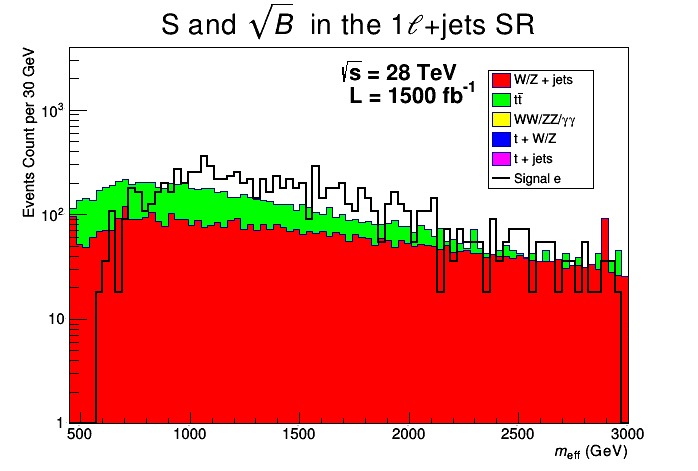}
   \includegraphics[width=0.40\textwidth]{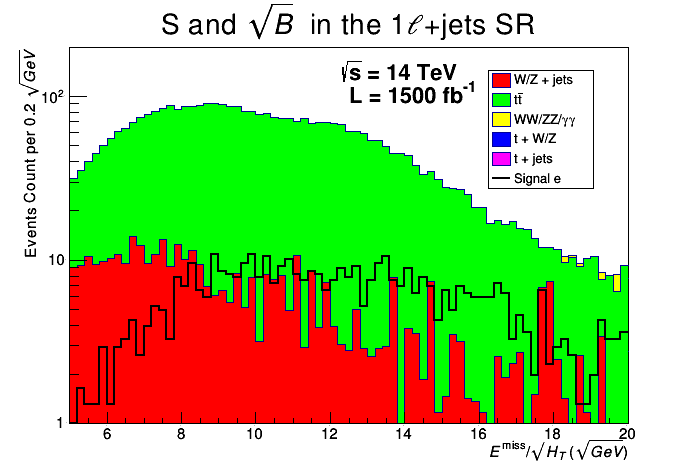}
   \includegraphics[width=0.40\textwidth]{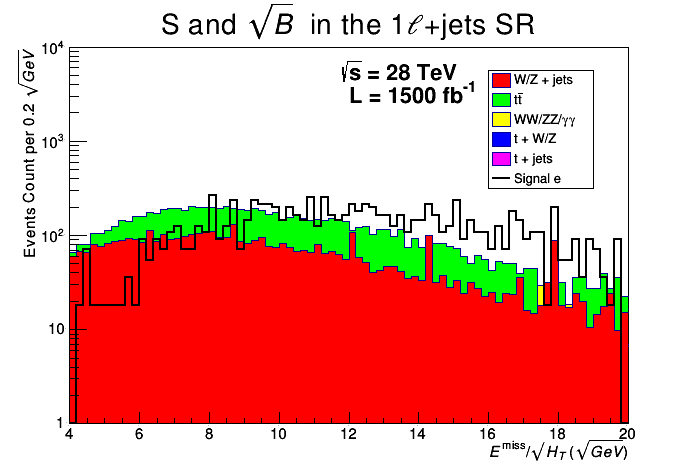}
    \caption{Distributions in the kinematic variable $m_{\rm eff}$ (top) and $E^{\rm miss}_T/\sqrt{H_T}$ (bottom) for benchmark point (e) at 14 TeV  (left) and 28 TeV (right) in the single lepton channel.}
	\label{fig3}
\end{figure}

\begin{figure}[H]
 \centering
 	\includegraphics[width=0.40\textwidth]{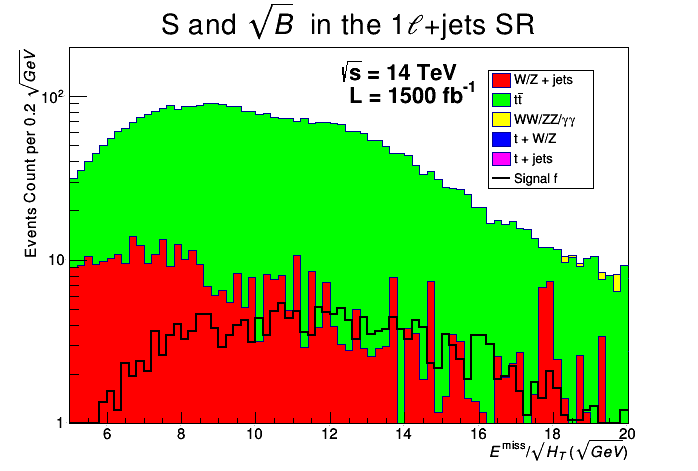}
   \includegraphics[width=0.40\textwidth]{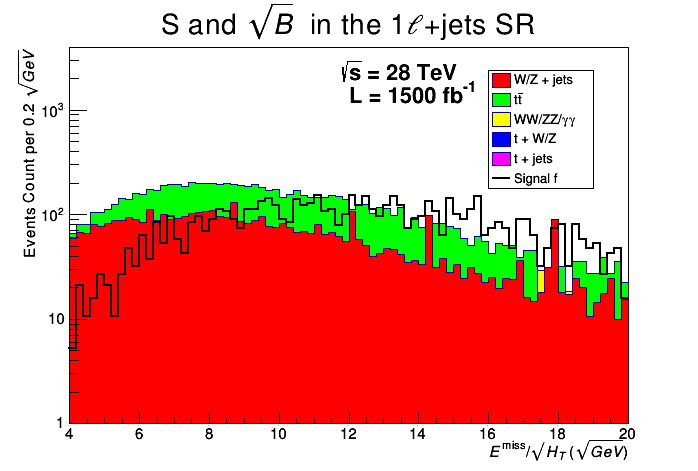}
   \includegraphics[width=0.40\textwidth]{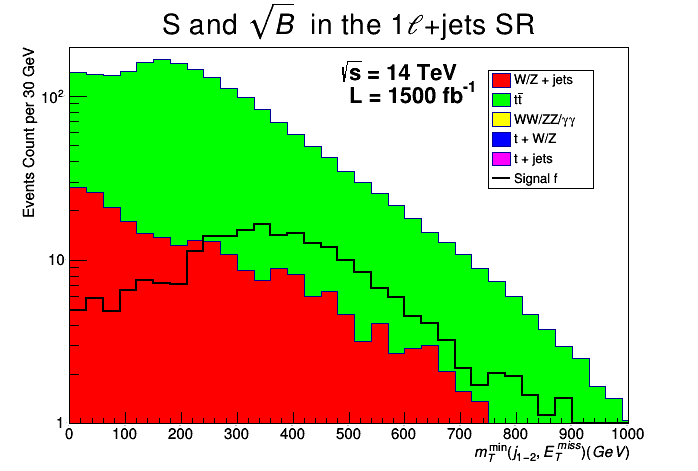}
   \includegraphics[width=0.40\textwidth]{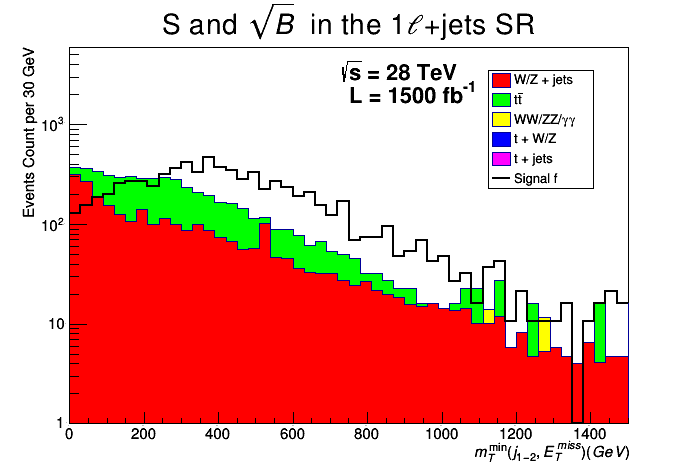}
    \caption{Distributions in the kinematic variable $E^{\rm miss}_{T}/\sqrt{H_T}$ (top) and $m^{\rm min}_{T}(j_{1-2},E^{\rm miss}_{T})$ (bottom) for benchmark point (f) at 14 TeV (left) and 28 TeV (right) in the single lepton channel.}
	\label{fig4}
\end{figure}

\begin{figure}[H]
 \centering
 	\includegraphics[width=0.40\textwidth]{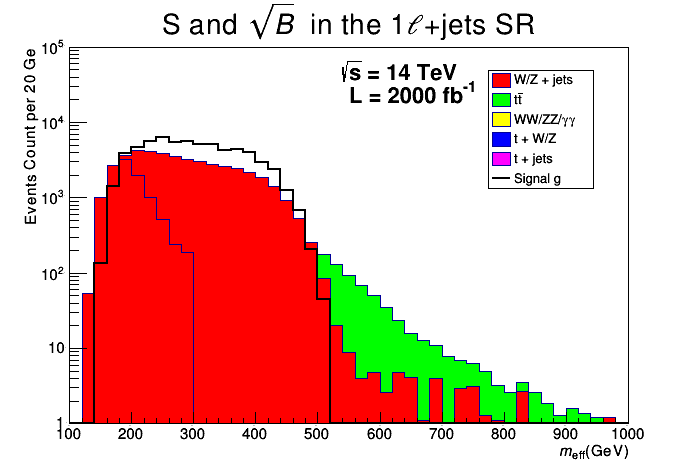}
   \includegraphics[width=0.40\textwidth]{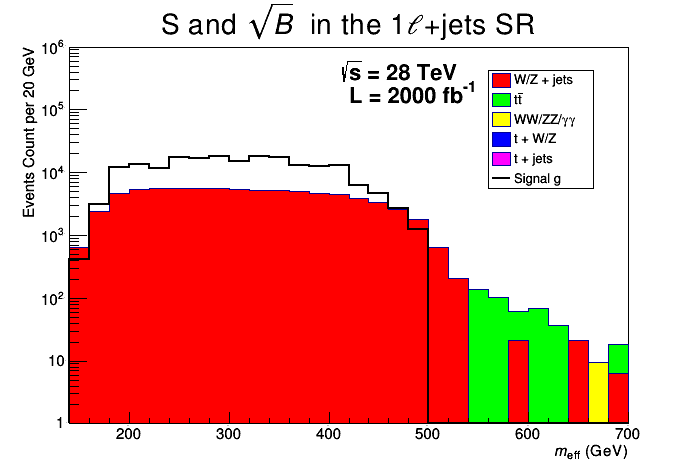}
   \includegraphics[width=0.40\textwidth]{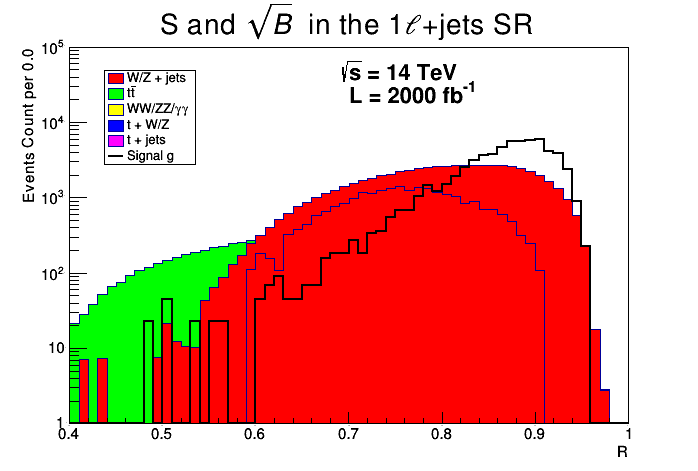}
   \includegraphics[width=0.40\textwidth]{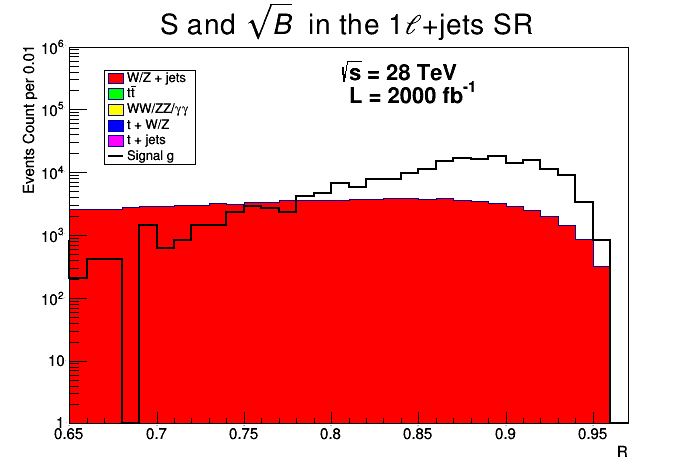}
    \caption{Distributions in the kinematic variable $m_{\rm eff}$ (top) and $R$ (bottom) for benchmark point (g) at 14 TeV (left) and 28 TeV (right) in the single lepton channel.}
	\label{fig5}
\end{figure}

\begin{figure}[H]
 \centering
 	\includegraphics[width=0.49\textwidth]{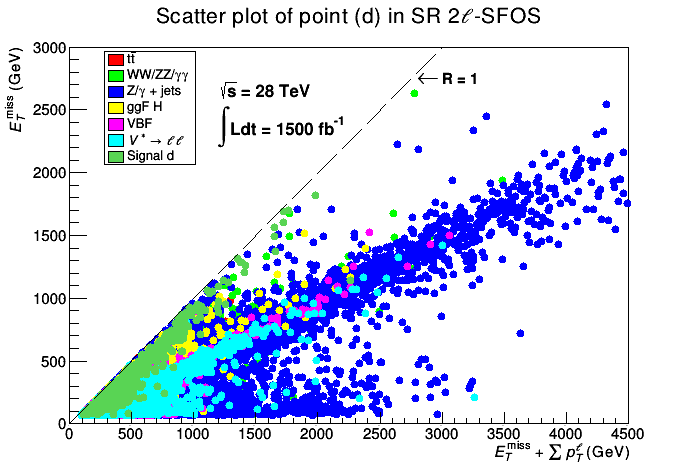}
   \includegraphics[width=0.49\textwidth]{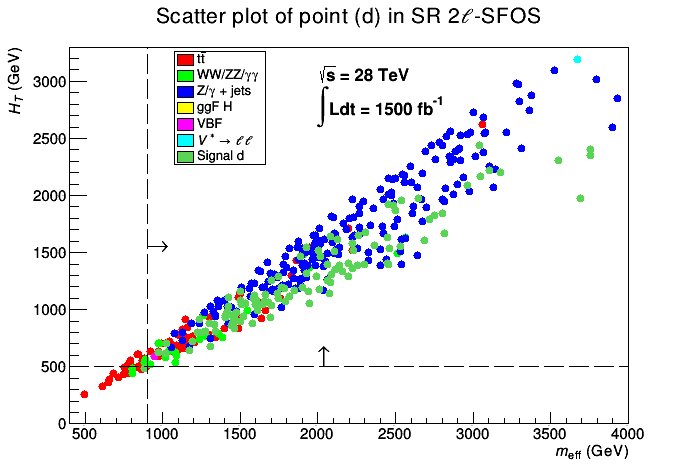}
   \includegraphics[width=0.49\textwidth]{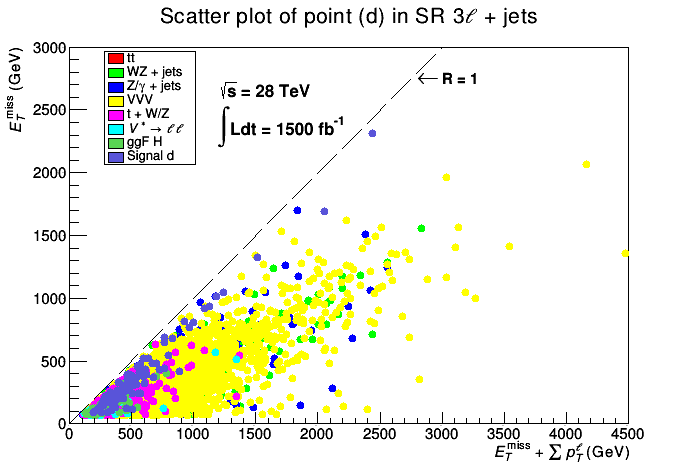}
    \caption{Scatter plots in the $(E^{\rm miss}_T+\sum p^{\ell}_T)$-$E^{\rm miss}_T$ plane for point (d) in the SR-$2\ell$-SFOS (top left panel) and in SR-$3\ell$ (bottom panel) at $\sqrt{s}=28$ TeV. Right top panel: scatter plot in the $m_{\rm eff}$-$H_T$ plane for the same point in SR-$2\ell$-SFOS.}
	\label{fig6}
\end{figure}

\end{document}
